\documentclass{elsart}

\usepackage{graphicx}

\usepackage{amssymb}

\begin{document}

\begin{frontmatter}

\title{Green's Function Measurements of Force Transmission in 2D Granular Materials}

\author[us]{Junfei Geng}, 
\author[france]{G. Reydellet},
\author[france]{E. Cl\'{e}ment},
\author[us]{R.~P. Behringer\corauthref{cor1}}
\ead{bob@phy.duke.edu}
\corauth[cor1]{Corresponding author.}

\address[us]{Center for Nonlinear and Complex Systems, Duke University,
Durham NC, 27708-0305, USA}
\address[france]{Universit\'{e} Pierre et Marie Curie, Paris 75231, France}

\begin{abstract}

We describe experiments that probe the response to a point force of 2D
granular systems under a variety of conditions.  Using photoelastic
particles to determine forces at the grain scale, we obtain ensembles
of responses for the following particle types, packing geometries and
conditions: monodisperse ordered hexagonal packings of disks,
bidisperse packings of disks with different amount of disorder, disks
packed in a regular rectangular lattice with different frictional
properties, packings of pentagonal particles, systems with forces
applied at an arbitrary angle at the surface, and systems prepared
with shear deformation, hence with texture or anisotropy. We
experimentally show that disorder, packing structure, friction and
texture significantly affect the average force response in granular
systems.  For packings with weak disorder, the mean forces propagate
primarily along lattice directions.  The width of the response along
these preferred directions grows with depth, increasingly so as the
disorder of the system grows.  Also, as the disorder increases, the
two propagation directions of the mean force merge into a single
direction.  The response function for the mean force in the most
strongly disordered system is quantitatively consistent with an
elastic description for forces applied nearly normally to a surface,
but this description is not as good for non-normal applied forces.
These observations are consistent with recent predictions of Bouchaud
et al. [Bouchaud et al., Euro. Phys. J. {\bf E4} 451 (2001); Socolar
et al., Euro. Phys. J. {\bf E7} 353 (2002)] and with the anisotropic
elasticity models of Goldenberg and Goldhirsch [Goldenberg {\&}
Goldhirsch, Phys. Rev. Lett. {\bf 89} 084302 (2002)].  At this time,
it is not possible to distinguish between these two models.  The data
do not support a diffusive picture, as in the q-model, and they are in
conflict with data by Rajchenbach [Da Silva {\&} Rajchenbach, Nature
{\bf 406} 708 (2000)] that indicate a parabolic response for a system
consisting of cuboidal blocks.  We also explore the spatial properties
of force chains in an anisotropic textured system created by a nearly
uniform shear.  This system is characterized by stress chains that are
strongly oriented along an angle of 45$^o$, corresponding to the
compressive direction of the shear deformation.  In this case, the
spatial correlation function for force has a range of only one
particle size in the direction transverse to the chains, and varies as
a power law in the direction of the chains, with an exponent of -0.81.
The response to forces is strongest along the direction of the force
chains, as expected.  Forces applied in other directions are
effectively refocused towards the strong force chain direction.

\end{abstract}

\end{frontmatter}


\section{Introduction}

\hspace{0.1in} Force propagation in granular materials is a fundamental, but
unresolved problem\cite{reviews,degennes_99} which has received much
recent 
attention\cite{reydellet_01,gengprl_01,rajchenbach,doubleY,socolar_02,goldhirsch,witten,witten_99,witten_00,mueggenburg_02}.
Several features of granular materials are responsible for the
complexity of the problem.  One of these is the fact that typical
materials do not exist in ordered states.  Here, order or disorder
involves several aspects.  The packing of grains is usually not in an
ordered lattice.  In addition, even packings with a high degree of
spatial packing order need not have order in the forces at the
particle contacts.  One cause of disorder is redundancy of contacts,
i.e. the fact that packings may have more contacts than are needed for
mechanical stability.  For example, in a hexagonal packing of ideal
disks, each disk except those at boundaries may have as many as six
contact points, while the conditions of force and torque balance (in
2D) require four contacts, i.e. two contacts located below the center
of gravity of each disk (hyperstatic equilibrium\cite{moukarzel}). In
real packings, it is often possible that each particle (even a
particle with low friction) can randomly lose several contacts without
destroying the stability of the lattice.  Conversely, even in a
packing of frictionless particles, there can be a substantial range of
forces at contacts, with a high degree of randomness. If friction is
introduced, non-normal forces are allowed, the number of degrees of
freedom increases, and effectively, the conditions for stability are
relaxed even further.  The frictional forces at grain contacts provide
an additional source of complexity: static frictional contacts are
history-dependent.  Hence, a seemingly ``ordered'' system from the
point of view of geometrical packing can contain disorder in the
contact forces, due to small shape and size variation of disks and,
more importantly, to the existence of
friction\cite{gengprl_01,duran,eloy,breton,gay_02}.  In addition, forces at
contacts are nonlinear, first, because forces vanish once a contact is
broken, and second, because in many cases, particles that are in
contact repel each other with normal forces that vary nonlinearly with
the inter-particle center of mass separation.  At the mesoscopic
level, even a nominally uniform applied external stress results in a
filamentary network of stress/force chains\cite{chains}, where a
modest fraction of the total number of grains carry the majority of
the force. We show an example of these chains in
Fig.~\ref{fig:shearcell}.

\hspace{0.1in} Repetition of experiments under identical macroscopic
conditions typically leads to substantially different stress chain
patterns each time.  This large variability under repetition suggests
that a statistical approach might be the most appropriate one.  This
approach might take the form of averaging a single realization over
large regions of space.  Alternatively, it might take the form of an
ensemble of measurements under identical macroscopic conditions.  The
assumed validity of the former is implicit in typical macroscopic
models of stress propagation for granular materials.  Here, we take
the second approach.  By determining an ensemble of responses at the
microscale, we not only determine average behavior, we also determine
the range of possible behavior and the probability of obtaining a
particular response to an applied stress or force.

\hspace{0.1in} A number of substantially different
models\cite{doubleY,socolar_02,goldhirsch,witten_99,continuum,coppersmith,osl,claudin98,cates_98}
exist to characterize force propagation in dense granular materials,
ranging from lattice\cite{socolar_02,coppersmith,claudin98} to
continuum\cite{continuum} descriptions.  The range of predictions is
underscored by noting that in the continuum case, various models
involve PDE's of totally different type.  For example, classical
elastoplastic models\cite{continuum}, are described by elliptic
equations below the plastic threshold (which is the region we consider
here) or hyperbolic equations above the plastic threshold.  The
continuum limit of the q-model of Coppersmith et
al.\cite{coppersmith}.  is a parabolic PDE.  The oriented stress
linearity (OSL) model\cite{osl} of Bouchaud et al. is based on a
hyperbolic PDE.  These authors have further explored force propagation
through lattice models which predict a wave equation for stress
propagation for ordered systems, a convection-diffusion equation for
weak disorder, and eventually, a transition to elliptic PDE's in the
presence of stronger disorder.

\hspace{0.1in} Very recently, two substantially different models have been proposed
to account for recent measurements\cite{reydellet_01,gengprl_01}.  One
is the force chain splitting model or double-Y model of Bouchaud et
al.\cite{doubleY,socolar_02} which is a Boltzmann equation for the
probability density of force chains with a given intensity and
orientation. In the presence of strong disorder and isotropic
``scattering'' of force chains, the authors derive stress equations
formally identical to those of classical elasticity. An alternative
model by Goldenberg and Goldhirsch\cite{goldhirsch} assumes that
nearest neighbors in a 2D packing of disks are coupled by
bi-directional or
uni-directional linear springs.  These authors propose that the
experimental results can be described using anisotropic elasticity,
leading to a PDE of elliptic type in the continuum limit. Very recently, there has been further exploration of elasticity models by Otto et al.~\cite{otto_03}.

\hspace{0.1in} Another important aspect of the problem concerns textures in a
granular system.  Texture refers to the distribution of contacts
between grains, and it is defined at the local scale, in terms of the
dyadic tensor formed from the components of the unit vectors between
the contacts experienced by particle and its center of mass
\cite{degennes_99,goddard,latzel}. Specifically, a fabric tensor
representing the distribution of contacts can be defined by the dyadic
product\cite{goddard,latzel}:
\begin{equation}
F_{xy} = < n_x^\alpha n_y^\alpha >.
\label{eq:fabric}
\end{equation}
Here, $\vec{n}^\alpha=n^{\alpha}_x \hat{x}+n^{\alpha}_y \hat{y}$ is
the unit vector from the center of mass of the particle to the
$\alpha$'th contact point, and the angle brackets represent an average
over all contact points on the particle. Both experiments and
numerical
simulations\cite{3dpile,2dpile,mueggenburg_02,matuttis,liffman} have
shown that the existence of non-isotropic textures due to different
deposition procedures of sandpiles or other packing procedures can
determine the way forces are transmitted and produce different stress
distributions. Among recent models, the force chain splitting model
emphasizes the need to incorporate a texture\cite{doubleY}.  The
spring model by Goldenberg and Goldhirsch\cite{goldhirsch} explores
the possibility of anisotropic elasticity associated with texture to
account for the features observed in experiments on ordered systems,
where forces tend to propagate along principal lattice directions.

\hspace{0.1in} A useful experimental tool for distinguishing among
these models is the response function for a localized
force\cite{degennes_99}. To the extent that this response is linear,
it corresponds to an experimental realization of the force Green's
function.  In previous work\cite{reydellet_01,gengprl_01}, we
investigated one of the simplest cases: response to a small force
applied normally at the boundary of 3D and 2D systems. We showed that
spatial ordering of the particles is a key factor: 2D ordered packings
respond strongly along the lattice directions, whereas disordered
packings show a broad elastic-like response both in 2D and 3D.  Other
recent experiments by Rajchenbach\cite{rajchenbach} involved a packing
of rectangular blocks, and the measured response was consistent with a
diffusive description.  The reasons for the disagreement between our
experiments and those of Rajchenbach remain unexplained, but may be
related to the differences in the particle types used in the two
different experiments.  Recently, Mueggenburg et
al.\cite{mueggenburg_02} reported measurements on ordered and
disordered 3D packings.  They found for disordered packings that there
was a broad central response to a point force, but the dependence of
the width on depth was not sufficiently well resolved to distinguish
between elastic vs. diffusive descriptions.  For ordered packings,
they found propagation along preferred directions.  However, the
nature of these directions and the amount of spreading of the response
depended crucially on the packing structure. FCC packings showed force
transmission along three well defined lines with moderate broadening
with depth.  By contrast, HCP packings showed substantially more
broadening with depth, and the preferred transmission was along cones.
These authors have interpreted their results in terms of the packing
geometry: for their FCC packings, they note the presence of direct
lines along which the forces could propagate, whereas in their HCP
packings, such paths did not exist.

\hspace{0.1in} This paper further explores through experiments the
issues of the local response of granular materials to small forces.
We consider the regime of limited deformation, so that to a reasonable
approximation, our results represent a true Green's function.  We
present information on the methods used and we explore a broad range
of systems.  In particular, we consider: 1) responses of monodisperse
disks in ordered hexagonal packings, 2) bidisperse systems with
different amount of disorder, 3) responses of systems of pentagonal
particles--systems that are primarily disordered, 4) responses of
rectangular packings of disks where we vary the inter-particle
friction, 5) responses to forces applied at arbitrary angles to the
surface, and 6) responses of a previously sheared, thus
textured/anisotropic, system.

\hspace{0.1in} The organization of this work is as follows. In Section 2, we
describe experimental procedures, methods and issues common to all
experiments. In Section 3, we describe a variety of experiments.
Finally, we draw conclusions in Section 4.

\section{Techniques, procedures and common features}

\hspace{0.1in} In this section, we describe procedures and analysis methods common to
all the experiments described here.  We also address the issue of
linearity in the force response.  An additional issue that is common
to all these experiments, which we discuss briefly in this section, is
the nature and importance of fluctuations.

\subsection{Experimental procedures}

\subsubsection{Experimental arrangements}

\hspace{0.1in} Photoelastic measurements\cite{heywood}, involving
stress-induced birefringence provide in 2D a unique opportunity to
obtain information about the internal structure of granular materials.
The experiments we describe below typically use a layer of
photoelastic grains consisting of either disks or pentagonal
particles. All the particles are cut from flat sheets of a
commercially available material (Measurements Group, Inc. material
PSM-4) with a Young's modulus of 4 MPa and a Poisson ratio between 0.4
and 0.5.

\begin{figure}[]
\center{\includegraphics[width=4.0in]{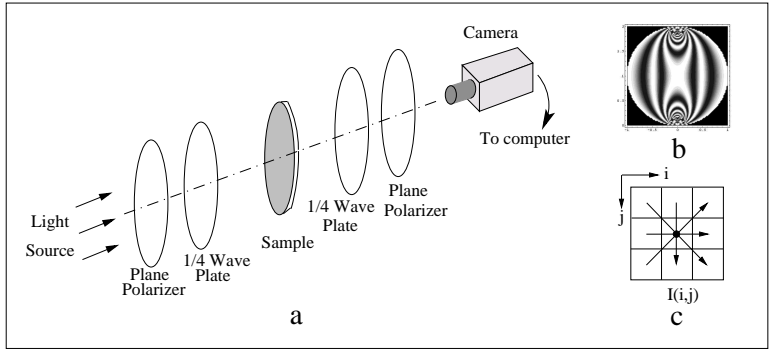}}
\caption{a) Schematic view of the circular polarimeter setup. In our
experiment each linear polarizer and quarter wave plate is combined
into a single sheet. b) Expected fringe pattern of a disk under
diametrical compression. c). Schematic drawing of how we calculate
$G^2$.}
\label{fig:setup}
\end{figure}

\hspace{0.1in} The grains are arranged in one of two configurations.
In one, they are contained between two transparent Plexiglas sheets;
in the second, the particles lean very gently against a glass plate
that is inclined by about $2^o$ from vertical, so that the grains are
supported, but with minimal friction with the supporting plate. In
either case, the assembly is placed between a pair of left- and
right-hand circular polarizers as shown in
Fig.~\ref{fig:setup}a. Light passes through this sandwich to produce a
polariscope intensity image.  We record these images with a digital
camera at a resolution of 640 $\times$ 480 pixels. For illumination,
we use a light box such as that used to read x-ray films, because this
provides a relatively homogeneous source.  When the photoelastic
grains are subjected to stresses, they become birefringent; the
resulting transmitted intensity is a measure of the applied stress, as
specified in more detail below.  Fig.~\ref{fig:setup}b shows a typical
intensity picture for a disk under diametrical compression observed
with polarizers.

\hspace{0.1in} When we use a vertical arrangement, the effect of hydrostatic head
must be removed. We note that most experiments were performed with
such an orientation because then the effects of friction with the
supporting plate are too small to be relevant.

\subsubsection{Force measurement}

\hspace{0.1in} A key issue is how to deduce forces on a particle, i.e. forces at a
grain scale.  When light travels through the particles along the
direction normal to the plane of the experiment, the emerging light
intensity, I, in a circular polariscope image is a function of the
stress in the plane of the disks at each position $(x,y)$, as in
Fig.~\ref{fig:setup}b.  Specifically, the local light intensity is
given by
\begin{equation}
I(x,y) = I_o \sin^2[(\sigma_2 -\sigma_1)t C /\lambda],
\end{equation}
where the $I_o$ is the incident light intensity, $\sigma_1$ and
$\sigma_2$ are the principle stresses at position $(x,y)$, $t$ is the
thickness of the sample, $C$ is the stress optic coefficient, and
$\lambda$ is the wavelength of the light.  In typical photoelastic
images of the particles, bands corresponding to different values of
$(\sigma_2 - \sigma_1)$ occur, where neighboring bright bands are
separated by a phase difference of $\pi$ in the argument of the sine
function above.

\hspace{0.1in} In general, the complete inverse problem that extracts vector forces
on a particle for a given photoelastic image is a formidable problem.
In these experiments, we use an empirical approach that allows us to
obtain force at the grain scale with reasonable accuracy and is much simpler than a complete calculation.  The basis for this process is the
fact that as the applied force at contact increases, the number of
fringes (black or white bands) also increases monotonically. We
exploit this fact to produce a force calibration in terms of a
quantity that we denote by $G^2$:

\begin{eqnarray} 
G^2 \equiv   |\nabla I|^2 & = &[ (\frac{I_{i-1,j}-I_{i+1,j}}{2})^2 + (\frac{I_{i,j-1}-I_{i,j+1}}{2})^2  +  \nonumber \\ 
& &(\frac{I_{i-1,j+1}-I_{i+1,j-1}}{2 \sqrt{2}})^2 +
(\frac{I_{i-1,j-1}-I_{i+1,j+1}}{2 \sqrt{2}})^2]/4,
\label{g_def}
\end{eqnarray}

where $I_{i,j}$ is the intensity at pixel $(i,j)$, as shown in
Fig.~\ref{fig:setup}c. The indices $i,j$ are the discrete replacement
of the continuous variables $x,y$.  Note that to avoid directional
preference, the vertical, horizontal, and both diagonal gradients are
squared and averaged with appropriate weights.  We first compute
$G^2(i,j)$ for each pixel $(i,j)$.  Then, for a particle or collection
of particles covered by $N$ pixels we calculate the average square
gradients: 

\begin{equation}
\langle G^2 \rangle= \frac {1}{N} \sum_{k=1}^N |\nabla I_k|^2.
\end{equation}

\begin{figure}[]
\center{\includegraphics[width=3.75in]{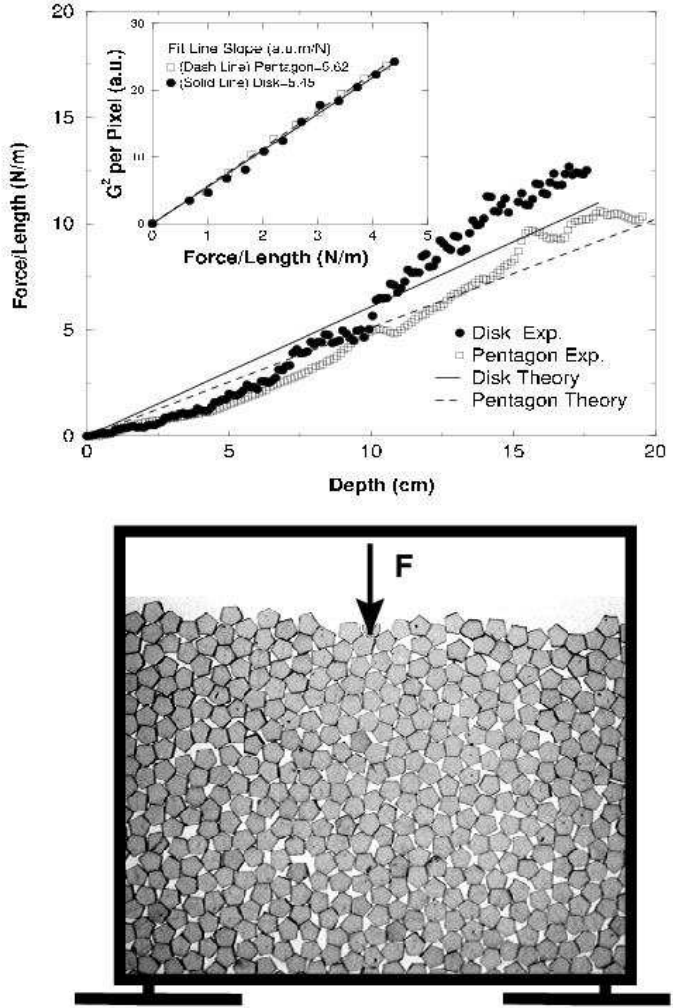}}
\caption{ a) Hydrostatic pressure due to gravitational force alone
versus depth determined from $G^2$. The expected slopes of the
stress-height curves are calculated from the known packing fraction
($\gamma=0.91$ for disks; $\gamma \simeq 0.75$ for pentagons). Inset
shows the multi-particle $G^{2}$ calibration by applying known loads
to the upper surface of the layer. b) Schematic of the experimental
apparatus with an overlay showing an image of an actual packing of
pentagonal particles. The drawing is not to scale.  The width is about
three times the height in the actual experiment, in order to avoid
boundary effects. F is the applied vertical load.}
\label{fig:cali}
\end{figure}

As the number of fringes increase, so does this average square
gradient. The method can be applied to calibrate the mean force on a
single particle or a larger assembly of particles.  We obtained
calibrations by either: (1) applying known forces to the boundary of a
small number of particles and at the same time measuring $\langle G^2
\rangle$, or (2) by applying various uniform loads to the upper
surface of a large rectangular sample (width larger than height to
avoid the Janssen effect) as shown in Fig.~\ref{fig:cali}b. We show
here a calibration curve, Fig.~\ref{fig:cali}(a) inset, using the
second method.

\hspace{0.1in} The validity of the $G^2$ calibration method is also 
tested by measuring the hydrostatic pressure vs. depth z. 
In a static system without external load, the hydrostatic 
pressure due to gravity is: $P=\gamma\rho g t z$, where 
$\rho$ is the density of the material from
which the particles are made, g is gravity and $\gamma$ is the packing
fraction of the particles. In the experiment, the density, $\rho$, is
1.15$\times$$10^{3}$$Kg\cdot m^{-3}$ and the typical packing fraction
$\gamma$ is $\sim$ 0.75 for pentagons and 0.91 for hexagonally packed
disks. The expected curves, based on the calibration of
Fig.~\ref{fig:cali}a-inset, and experimental hydrostatic head data
obtained from the $G^2$ method, shown in Fig.~\ref{fig:cali}a, agree
well. Thus, a simple interpretation of the $G^2$ method is that it
represents the local pressure (i.e. the trace of the stress tensor).
These calibrations are effective until the forces are so large that it
is no longer possible to resolve the fringes on a particle clearly, a
condition that does not occur in these studies.  They are also limited
at low forces, since the gradients in that case become weak.

\subsubsection{Procedures}

\hspace{0.1in} A typical procedure was as follows.  The particles were first placed
in the apparatus.  We then obtained a sequence of images.  The first
image, made in the absence of the applied load, yielded the particle
locations.  This image was taken without the polarizers in place, and
from it, we extracted particle positions and contacts by image
analysis.  A second image
with polarizers in place but no applied load provided the background
photoelastic image.  We then measured the system point-force response
by placing a known weight carefully on top of one particle at the
surface or by pushing on one grain with a high precision digital force
gauge (model DPS-110 from Imada Inc.).  With the local applied force
in place, we obtained a second photoelastic image.  We then removed the local
force and obtained one last image without polarizers to ascertain if
there had been any particle movement.

\hspace{0.1in} With the exception of packings of bidisperse disks, we did not use
trials in which there were changes in the particle packing after the
local force was removed.  In the case of bidisperse disk packings,
some small movement of the particles at the surface typically occurred
no matter how weak the applied force.  However, this motion was
limited to particles very near the surface and the location of the
applied force.

\hspace{0.1in} By computing $G^2$ at the pixel
scale for each image and subtracting the background from the response
with load, we obtained the stress difference between successive images
of $G^2$, containing only the response from the point perturbation. We
refer to this difference as $\Delta G^2$, or as $G^2$ in the cases
where no confusion is caused.

\subsection{A statistical approach} 

\hspace{0.1in} In all cases, the responses differ significantly from
realization to realization, a feature that was also considered in
recent theoretical work\cite{claudin98}. This is true even for the
case of nearly regular grain packings, since the frictional forces at
the contacts are determined by the microscopic details of the
preparation history, something that in general is not known.  Hence,
it is necessary to develop an ensemble of measurements under identical
macroscopic control conditions in order to extract the mean behavior.
In order to obtain such ensembles, we repeated measurements on a given
system for many different rearrangements of the particles, typically
50 times for each set of data. Between runs, the system was either
``stirred'' using a rod or ``massaged'' using hand to rearrange the
particles. For the regularly packed lattices, the goal was to rearrange the forces at the contacts
without generally changing the positional order of the particles. For the disordered lattices, the positional order was also changed. To
make sure that our measurements were statistically significant, we
divided 50 measurements into two groups, each consists 25
measurements, and verified that the averaged responses for two groups
were consistent.

\hspace{0.1in} For responses in each realization, we denote the stress at the pixel
scale for a given realization, $n$, as $G^2(x,y,n)$.  To obtain the
ensemble mean response, we average in two ways.  First, we compute the average
of $G^2(x,y,n)$ over $n$.  As noted above, we then carry out a
coarse-graining at the scale of a single particle, since variations in
$G^2$ below the particle size are not meaningful here.  The result is
denoted by $\,$ $\overline{G^2(x,y)}$.

\begin{figure}[t]
\center{\includegraphics[width=3.75in]{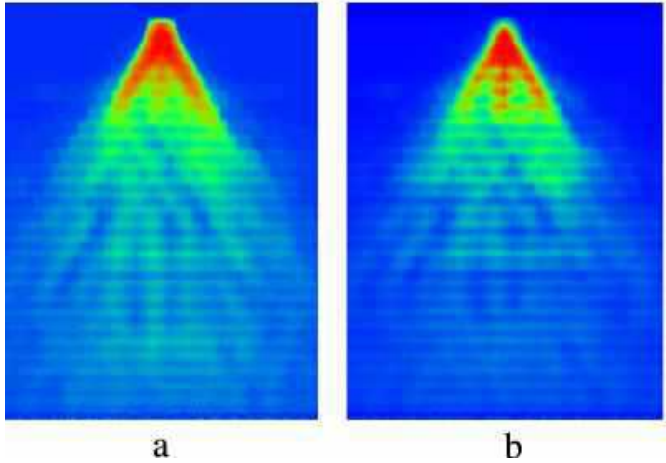}}
\caption{The mean response, (a), and the standard deviation, (b), of $G^{2}$ for a hexagonal packing of disks. The standard deviation image has a similar shape to the mean image.}
\label{fig:fluctuations}
\end{figure}

\begin{figure}[]
\center{\includegraphics[width=3.75in]{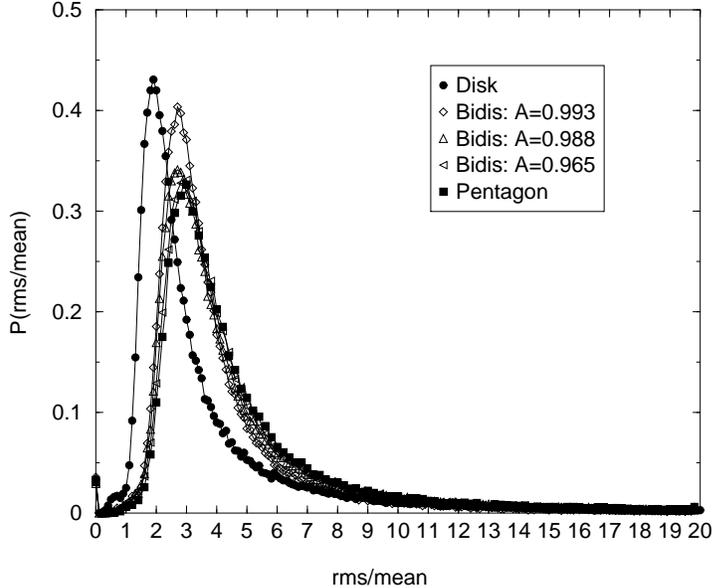}}
\caption{The distributions of the ratio of standard deviation to the
mean $G^{2}$ for all points in the system. Each curve represents a
different system, as indicated in the inset.}
\label{fig:fluct-quant}
\end{figure}

\hspace{0.1in} The ensemble contains important information concerning the range
and probability of results that might be encountered
at any position $(x,y)$.  To
characterize the fluctuations from one realization to another, we
calculate the standard deviation of the stress for each position:
$rms(x,y)=\sqrt{Var}$, where
\begin{equation}
Var=\frac{1}{N-1}\sum^{N}_{n=1}(G^2(x,y,n)-\overline{G^2(x,y)})^2.
\end{equation}
As an example, we show the $rms(x,y)$ for a hexagonal disk packing in
Fig.~\ref{fig:fluctuations}b, using a greyscale representation.  Here,
brighter regions represent larger values of the $rms$. We contrast the
greyscale image for the fluctuations to the mean response in
Fig.~\ref{fig:fluctuations}a. The $rms$ image clearly has a similar
shape to the mean image, and the similarity in these patterns suggests
that there is a simple point-wise relation between these two
quantities.  We explore this by determining the distribution, without
regard to position, of the ratio $rms(x,y)/\overline{G^2(x,y)}$ for
all points in the system.  The data for the hexagonal packings of
monodisperse disks are given by the solid circles in
Fig.~\ref{fig:fluct-quant}. The distribution has a well defined peak
at around 2, i.e. the most probable occurrence is that the $rms$ is
twice the mean.  The rest of the curves in Fig.~\ref{fig:fluct-quant}
are similar plots for bidisperse systems of disks with different
amount of disorder, and for a system of pentagons, systems which we
will discuss below.  Briefly, the amount of spatial disorder increases
as we change the system from monodisperse disks to bidisperse disks
and finally to pentagons.  As the amount of disorder increases, the
peak in the distributions of $rms/\overline{G^2(x,y)}$ shifts to
larger values, and the distributions become wider.

\subsection{Linearity of the Response}

\hspace{0.1in} Additional issues that are of considerable importance in all the
measurements are reversibility and linearity.  The first refers to the
fact that the particles return to their unperturbed state after the
applied local force is removed.  The second concerns the functional
relationship between the size of the applied force and the response at
a given point.  With the exception of bidisperse systems of disks (as noted
above) for
the measurements reported here, deformations were completely reversible
for small forces.  For bidisperse packings, the process of
adding and removing a point force was reversible except for
a few of particles
on the surface and near the applied force.  The vast majority of
the particles, those in the bulk of the
sample were undisturbed.

\begin{figure}[tbh]
\center{\includegraphics[width=5.0in]{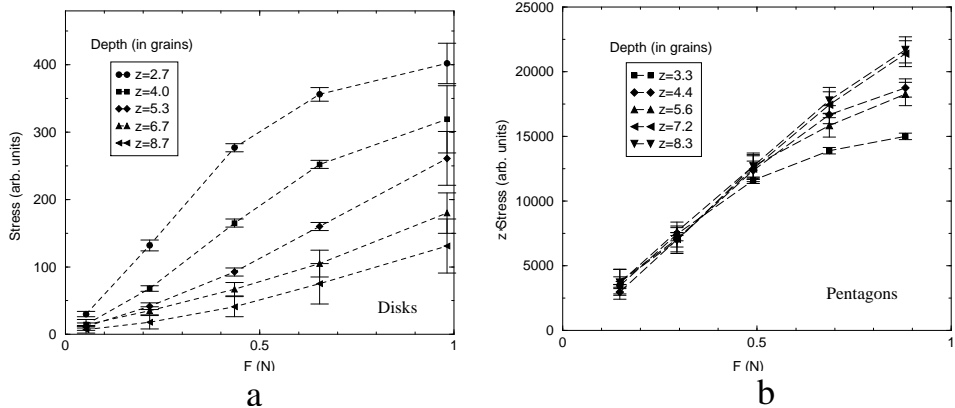}}
\caption{Linearity test: a) Measured peak stress vs. applied force at
different depths along the principal lattice directions in a
hexagonally packed monodisperse disk system. b). Measured stress
multiplied by the corresponding depth vs. applied force for different
depths in a pentagonal system.  (See Appendix)}
\label{fig:lin_test}
\end{figure}

\hspace{0.1in} However, reversibility does not necessarily mean linearity.  We have
carried out systematic tests of the linearity of the response, as
shown in Fig.~\ref{fig:lin_test}.  These data are discussed
in more detail in the appendix, where we place them in the
context of appropriate models.  For both ordered triangular disk
packings and disordered pentagonal packings, when the applied force is
below about 0.5 N, there is a reasonable linearity within the error
bars. In order to optimize the signal to noise ratio, yet avoid
nonlinear effects, we usually chose a working force close to the upper
bound of this linear region.

\section{Experimental results}

\hspace{0.1in} In this section, we describe four classes of experiments that address
various factors affecting the force propagation in granular systems,
including disorder, packing structure, friction, direction of applied
forces, and textures.

\subsection{Role of packing disorder: Responses of bidisperse systems}

\hspace{0.1in} As noted, disorder at particle contacts may arise from at least two
sources.  One is the presence of geometrical disorder in the packing.
The other is the random disorder in the contact forces, due for
instance, to frictional indeterminacy.  We first consider the effects
of geometrical disorder on the packing.

\hspace{0.1in} The first way that we did so was by determining the
force response for bidisperse systems with varying amounts of packing
disorder.  We modified the amount of disorder in a controlled way as
follows: We prepared each sample by mixing about 500 small and 500
large disks in a container, so that $n_1 \simeq n_2\sim0.5$.  We then
randomly chose one particle to add to the upper surface of the sample
until the full amount of particles was in place. The exact values of
$n_1$ and $n_2$ were determined later from images showing particle
configurations using particle identification software mentioned above.
The experiments on bidisperse disks were all performed in a nearly
vertical plane\cite{gengprl_01}.

\hspace{0.1in} It is important to characterize the amount of disorder in the samples.
To this end, we have pursued two approaches.  The first method
involves a parameterization of the width of the disk radius
distribution $w(a)$ through the parameter ${\mathcal{A}} = \langle a
\rangle^2/\langle a^2 \rangle$, as proposed by Luding
et al.\cite{gengprl_01,luding}
This parameter has proved useful in
characterizing the bidispersity of granular systems in the kinetic
regime\cite{luding}.  For our purposes, it is also useful, because
it provides a relatively precise way to label bidisperse systems.
Specifically,
the moments of the distribution $\langle a^m \rangle$ are given by
$\langle a^m \rangle = \int w(a) a^m da / \int w(a) da$.  ${\mathcal{A}}=1$
corresponds to perfect order in a monodisperse situation, and the
deviation from unity is proportional to the degree of poly-dispersity
or disorder in the system.
In a bidisperse system, with respective radii
of smaller and larger particles $a_1$ and $a_2$, and corresponding
particle numbers $N_1$ and $N_2$, the parameter ${\mathcal{A}}$ is:
\begin{equation}
{\mathcal{A}} = \frac{\langle a \rangle ^2}{\langle a^2 \rangle}
         = \frac{[n_1+(1-n_1)/R]^2}{n_1+(1-n_1)/R^2}.
\end{equation}
Here, the size ratio is $R = a_1/a_2$, the number fractions are
$n_i=N_i/N$ ($i$=1,\,2) and the total number of particles is
$N=N_1+N_2$.  

\hspace{0.1in} Besides the ${\mathcal{A}}=1$ case for ordered monodisperse packing, we
have used disks with 3 different diameters (0.597 cm, 0.744 cm and
0.876 cm) and obtained bidisperse systems with three different ${\mathcal{
A}}$ values, i.e. ${\mathcal{A}}=0.993$, $0.988$ and $0.965$, as shown in Table 1.
\\
\\
\centerline{\begin{tabular}{|c|c|c|c|c|} \hline
\emph{System Type} & \emph{Disk Diameters} & \emph{R} & \emph{$n_1$} & {$\mathcal{A}$} \\ \hline
\hspace{0.1in}monodisperse\hspace{0.1in} & 0.744 cm &/ & / & 1 \\ \hline
\hspace{0.1in}bidisperse 1&0.744, 0.876 cm &\hspace{0.1in} 0.849\hspace{0.1in} &\hspace{0.1in} 0.590\hspace{0.1in} &\hspace{0.1in} 0.993\hspace{0.1in} \\ \hline
\hspace{0.1in}bidisperse 2& 0.597, 0.744 cm & 0.802 & 0.550 & 0.988 \\ \hline
\hspace{0.1in}bidisperse 3& 0.597, 0.876 cm & 0.682 & 0.520 & 0.965 \\ \hline
\end{tabular}}
\vspace{0.1in}
Table 1: Experimental control parameters for the ordered monodisperse system and three bidisperse systems, where $R$ is the size ratio between small and large disks, $n_1$ the number fraction of the small disk, and ${\mathcal{A}}$ the disorder parameter.
\\

\begin{figure}[b]
\center{\includegraphics[width=5.0in]{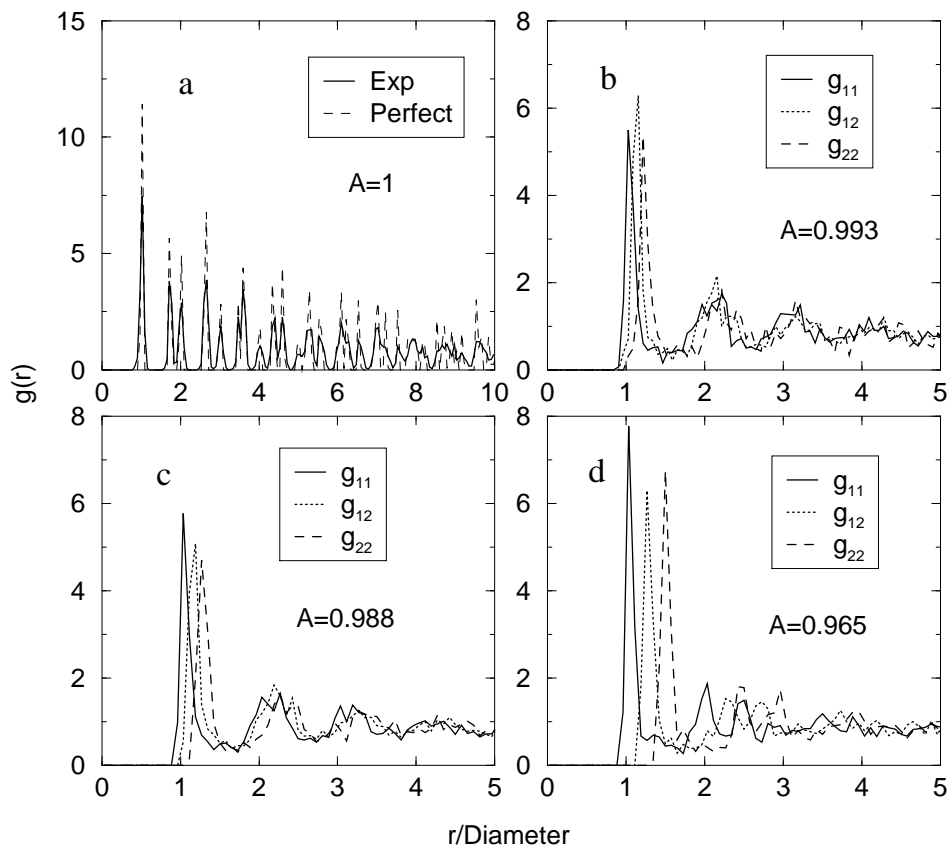}}
\caption{Particle-particle autocorrelation functions for a)
hexagonally packed monodisperse disks, b), c) and d) three bidisperse
systems with b) ${\mathcal{A}}=0.993$, c) ${\mathcal{A}}=0.988$ and d) ${\mathcal{A}}=0.965$, respectively. Distances are normalized by the diameter of
the smaller particles in each system. In a), the dashed line is
calculated from a perfect triangular lattice of comparable size to the
experiment, and the solid line is from the experiments. In b), c) and
d), $g_{11}$ and $g_{22}$ are correlation functions for the same
species of particles and $g_{12}$ are for different species. (see
Table I.)}
\label{fig:correl_bi_combo}
\end{figure}

\hspace{0.1in} An alternative method to quantify the disorder of the system is to
calculate the particle-particle positional autocorrelation
function or radial distribution function. The autocorrelation function is defined as \cite{luding},
\begin{equation}
g(r) =
 \frac{2A}{N(N-1)}\frac{1}{A_r}\sum_{i=1}^{N}\sum_{j=1}^{i-1}\theta(r_{ij}-r)\theta(r+\Delta
 r-r_{ij}), 
\end{equation} 
where $A$ is the area of the system, N the number of particles,
$A_r=\pi (2r+\Delta r)\Delta r$ the area of a ring between $r$ to
$r+\Delta r$, and $r_{ij}$ the distance between particle $i$ and $j$.
The two $\theta$ functions select all particle pairs with distances
between $r$ and $r+\Delta r$. For bidisperse systems, we calculate the
autocorrelation function between the same species and between
different species. When calculating the autocorrelation function for
different species, the weight $N(N-1)/2$ in the above equation must be
changed to $N_1N_2$ and the indices $i$ and $j$ run from $1$ to $N_1$
and $N_2$ respectively, in order to account for all pairs of different
kinds.

\hspace{0.1in} In Fig.~\ref{fig:correl_bi_combo}, we show correlation functions for
the monodisperse and three bidisperse disk systems that we have
investigated. In Fig.~\ref{fig:correl_bi_combo}a, we contrast the
correlation functions calculated from an ideal triangular disk lattice
and from an experimentally obtained triangular lattice of monodisperse
disks.  In the latter case, the broadening of the peaks indicates some
irregularity in the packing. However, both correlation functions show
a long range order with peaks at 1, $\sqrt{3}$, 2, ..., as expected for a
triangular lattice. The decay in the correlation functions is due to
the finite size of the system, and has a characteristic length of
$\sim 8$ particle diameters. In Fig.~\ref{fig:correl_bi_combo}b,$\;$c
and d, we show experimental correlation functions between the same
species and between different disk species for bidisperse systems. In
contrast to the monodisperse case, we see that over a distance of
several disk diameters, the correlations of bidisperse systems
decrease very quickly to the background value of 1, and the peaks
corresponding to the second and third coordination shell broaden and
merge with each other, indicating increasing disorder as the control
parameter ${\mathcal{A}}$ decreases.  The correlation lengths, $L$, for
bidisperse systems are $3.86$, $3.83$ and $3.72$, measured in disk
diameters, for systems of ${\mathcal{A}}=0.993$, $0.988$ and $0.965$,
respectively, which are calculated according to $L^2 = \int
r^2g(r)d\vec{r} /\int g(r)d\vec{r}$\cite{ziman}.  Interestingly, the
correlation function changes only slightly for these various packings,
even though the measured response changes significantly.

\begin{figure}[b]
\center{\includegraphics[width=5.0in]{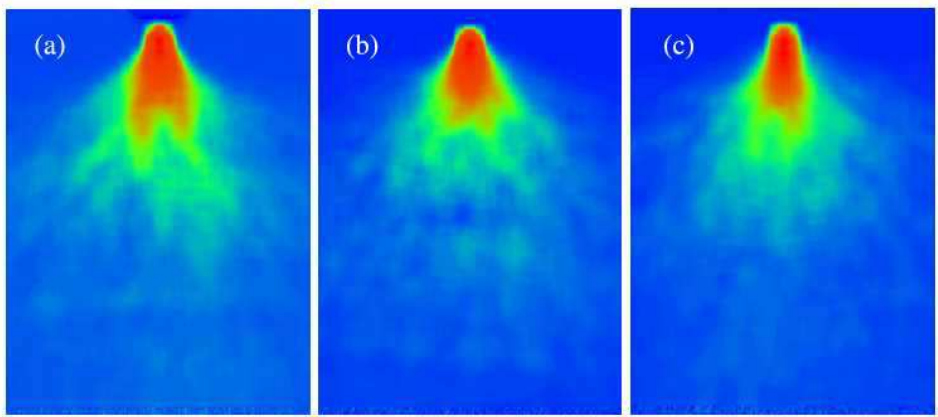}}
\caption{Mean response for 50 trials of a 50 g point force for
bidisperse systems of disks with different amounts of disorder, (a)
${\mathcal{A}}=0.993$ (b) ${\mathcal{A}}=0.988$ and (c) ${\mathcal{A}}=0.965$. The size of each image is $300 \times 400 $ pixels (about $18.0 \times
13.5$ cm).}
\label{fig:bidis_pics}
\end{figure}

\begin{figure}[p]
\center{\includegraphics[width=3.2in]{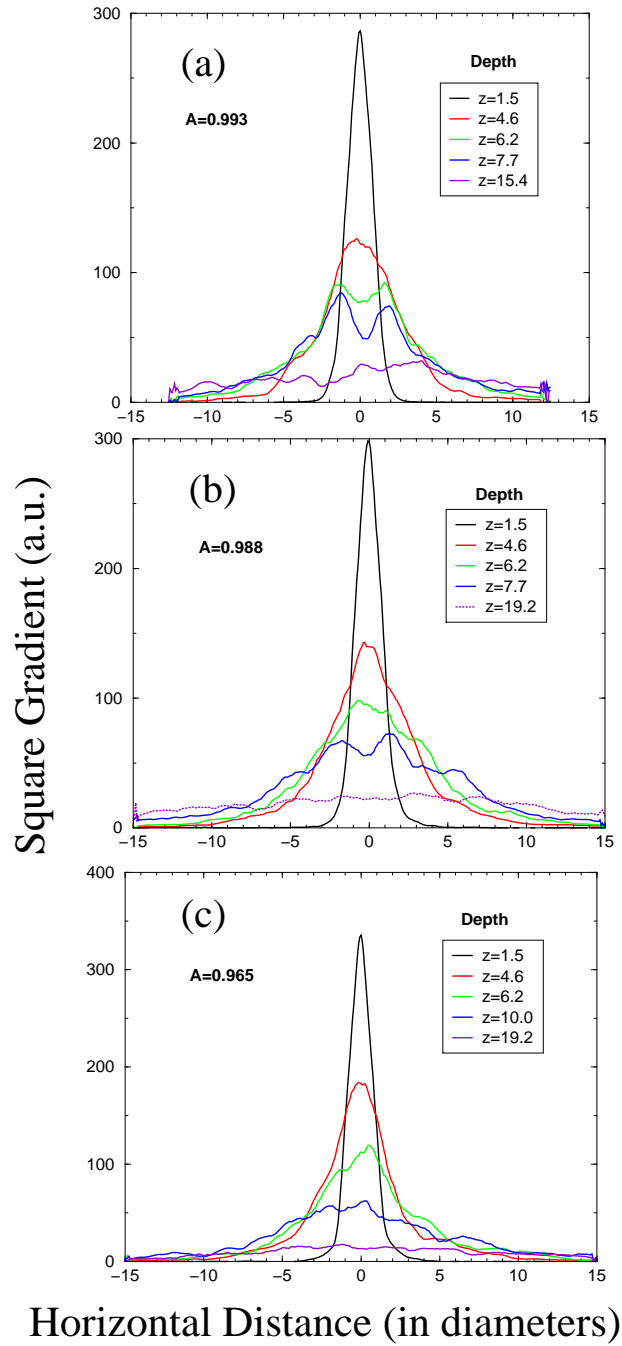}}
\caption{Photoelastic response, $G^2$, to a point force vs. horizontal
distance $x$ at various depths $z$ from the source for bidisperse disk
systems with a) ${\mathcal{A}}=0.993$, b) ${\mathcal{A}}=0.988$ and c) ${\mathcal{
A}}=0.965$, respectively. The horizontal distance and depth in the
plots are in units of the smallest relevant disk diameters.  The
diameters of three different sized disks are 0.597 cm, 0.744 cm and
0.806 cm. }
\label{fig:bidis_data}
\end{figure}

\hspace{0.1in} We now turn to the experimental results for bidisperse arrays.
By increasing the amount of disorder in the system,
we observed responses that change from a two-peak
structure to a response very similar to that of our most disordered
system, namely a system of pentagons.  In Fig.~\ref{fig:bidis_pics},
we give a greyscale representation of the average response to a point
force for the three bidisperse systems with ${\mathcal{A}}=0.993$, $0.988$ and
$0.965$.  In Fig.~\ref{fig:bidis_data}, we present the
same results by showing the response along a series of horizontal
lines at a number of depths, $z$, measured from the upper surface.  For the
largest ${\mathcal{A}}$, ${\mathcal{A}}=0.993$, Fig.~\ref{fig:bidis_pics}a and
Fig.~\ref{fig:bidis_data}a show responses with two-peak features that
resemble the response structure for ordered monodisperse disks.
However, with decreasing ${\mathcal{A}}$, this feature becomes
progressively weaker. In Fig.~\ref{fig:bidis_pics}c and
Fig.~\ref{fig:bidis_data}c, it has completely disappeared, and the
response is similar to that of system of pentagonal
particles\cite{gengprl_01}  (See also Figs ~\ref{fig:oblique_pent_pics} and ~\ref{fig:oblique_pent_data} below). The change from a two-peak to a one-peak
structure presents clear evidence of the important role of disorder in
force responses.

\subsection{Response for rectangular lattices of disks}

\hspace{0.1in} In the experiments of this subsection, we further consider how packing
structure and friction affect the average responses for rectangular
lattices.  The use of rectangular lattices tends to reduce the
randomness of forces at contacts, since all contacts are now essential
for stability of the packing.  That is, the number of
contacts between disks is minimal and the contact network is well
defined in the sense that the force at every contact is nonzero. We
emphasize, however, that randomness in contact forces still exists,
due to friction.

\begin{figure}[t]
\center{\includegraphics[width=3.75in]{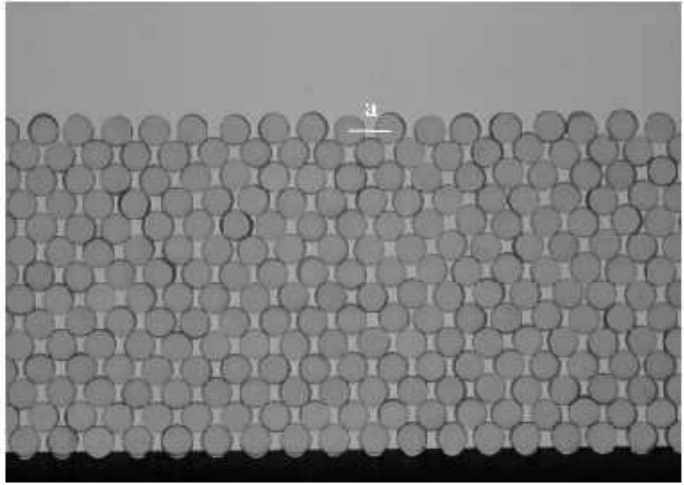}}
\caption{A rectangular packing of disks with a horizontal lattice
constant a=1.27d, where d is the disk diameter.}
\label{fig:rect}
\end{figure}

\hspace{0.1in} A typical rectangular monodisperse packing is shown in
Fig.~\ref{fig:rect}.  To construct this packing, disks on the bottom
layer were supported by a template consisting of equally spaced
grooves with a center-to-center spacing of 1.27 disk diameters.  The
system size was $\sim 85$ particles wide and $\sim 15$ particles
high. Since a rectangular lattice has less contacts than a triangular
lattice, it is also less stable than a triangular
lattice. Consequently, it was more difficult to build tall layers, and
once built, a layer could not support as large forces as in the case
of triangular packings. Therefore, in this set of experiments, we used
a 20 gram force for the probe.

\hspace{0.1in} We note that the naturally occurring surfaces of the disks were
relatively frictional, with a static friction coefficient $\mu$ close
to $0.94$. We estimated $\mu$ by placing two disks (glued together
side-by-side so that they could not roll) on a slope of same material
from which the disks are made, tilting the slope and then recording
the angle when the particles start to slip. In separate experiments,
we wrapped each disk with Teflon tape, thus reducing the friction
coefficients to about $\mu=0.48$. This allowed us to investigate the
role of disorder associated with friction in the force response.

\begin{figure}[t]
\center{\includegraphics[width=5in]{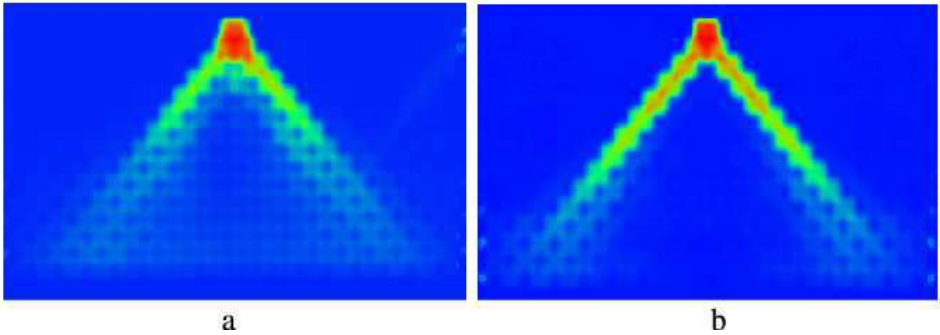}}
\caption{Mean response for 50 trials of a 20 g point force for
rectangular packings of disks: a) relatively frictional disks with a
coefficient of friction, $\mu$, close to 0.94, and b) less frictional
(Teflon wrapped) particles with $\mu=0.48$.}
\label{fig:resp_rect_pics}
\end{figure}

\begin{figure}[t]
\center{\includegraphics[width=5in]{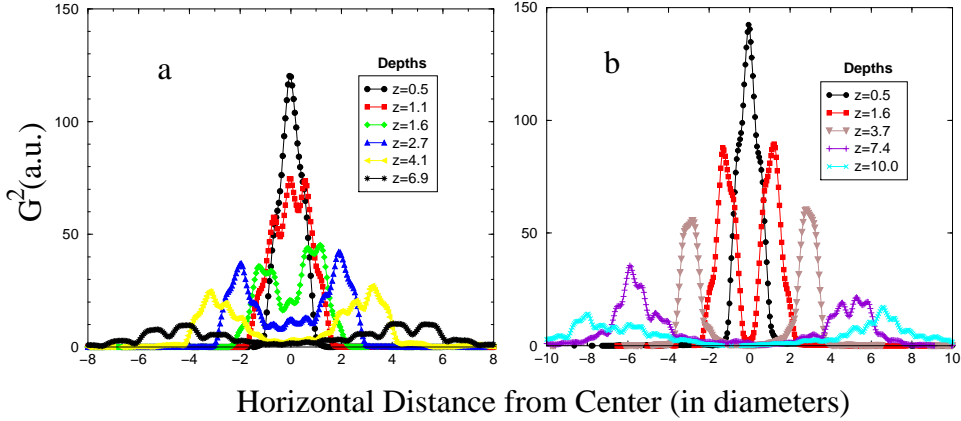}}
\caption{Photoelastic response $G^2$ to a point force, vs. horizontal
distance, $x$, at various depths, z, from the source for rectangular
packings of disks with different coefficients of friction: a)
$\mu=0.94$, and b) $\mu=0.48$.}
\label{fig:resp_rect_data}
\end{figure}

\begin{figure}[t]
\center{\includegraphics[width=5in]{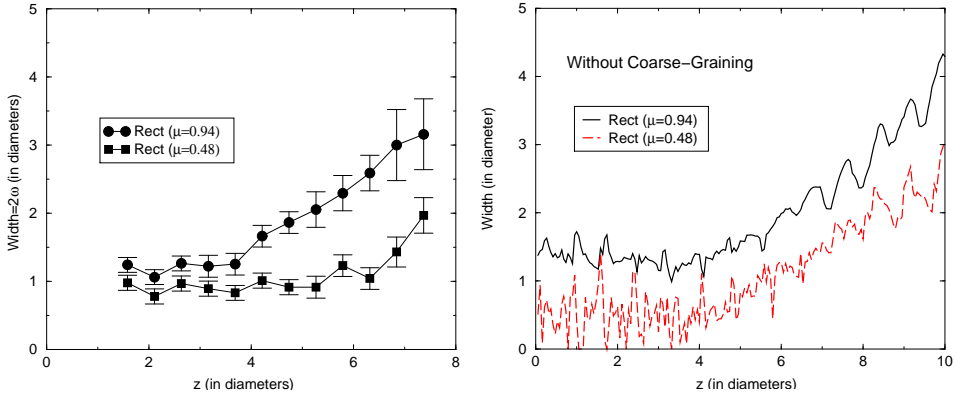}}
\caption{Width of peaks v.s. depth for rectangular packings of disks
with different coefficients of friction.  Data on the left show
results with the usual coarse-graining average over one diameter.
Data on the right is for the same measurements but without coarse
graining.}
\label{fig:rect_width}
\end{figure}

\hspace{0.1in} In Fig.~\ref{fig:resp_rect_pics}a-b, we show the grey-scale average
response pictures for the rectangular lattice systems with large and
small friction coefficients, respectively. In
Fig.~\ref{fig:resp_rect_data}, we show quantitative data at several
depths for both systems. In both cases, the responses propagate along
the lattice directions. The measured value of the angle between the
two propagation directions is $\sim 79^o$, which corresponds well with
the rectangular lattice structure.  This is qualitatively similar to
the results for a triangular lattice\cite{gengprl_01}, where the angle
between the two preferred directions was $60^o$. For both the higher
friction rectangular packing and the hexagonal packing, the peaks in
the response broadened relatively rapidly with depth. When the
friction was decreased (Teflon wrapped particles) the peaks remained
significantly sharper with depth, as shown in
Fig.~\ref{fig:rect_width}. Here, the width $w$ is obtained by fitting
a Gaussian curve $F=F_0e^{-\frac{(x-x_o)^2}{2w^2}}$ to each peak at a
given depth. Widths extracted by other means, for example, measuring
the width at half maximum height, give similar results. Note that for
our usual processing, in which we coarse-grain at the scale of one
particle diameter, the smallest peak width is one particle diameter,
as seen in the left part of Fig.~\ref{fig:rect_width}.  It is
interesting in this case to examine the width of the peaks for the
data without coarse graining, as shown in the right side of the
figure.  These data suggest that in a perfectly ordered and
frictionless system, the force response would be perfectly sharp force
chains.

\subsection{Comparison to Models}

\begin{figure}[t]
\center{\includegraphics[width=3.75in]{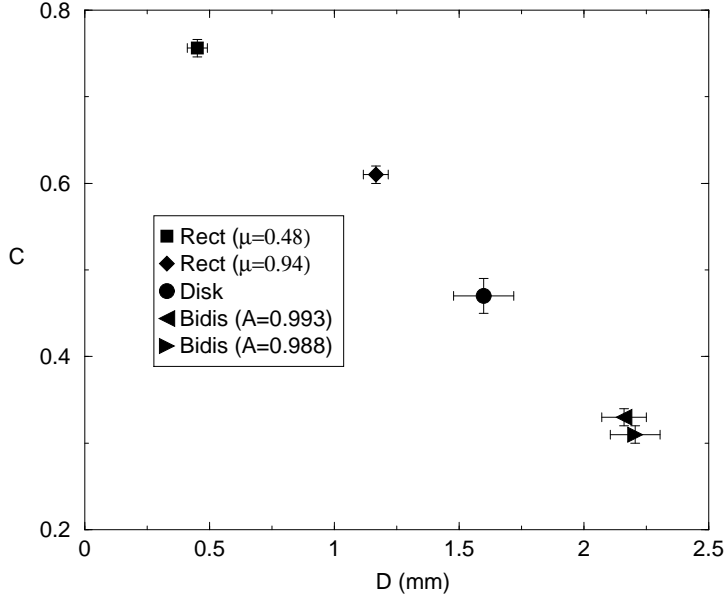}}
\caption{The coefficients of c and D extracted by fitting the mean
responses from different systems to the CD equation. Each point in the
graph corresponds to a different system. See the text for more
detail.}
\label{fig:CD_fit}
\end{figure}

\begin{figure}[t]
\center{\includegraphics[width=4in]{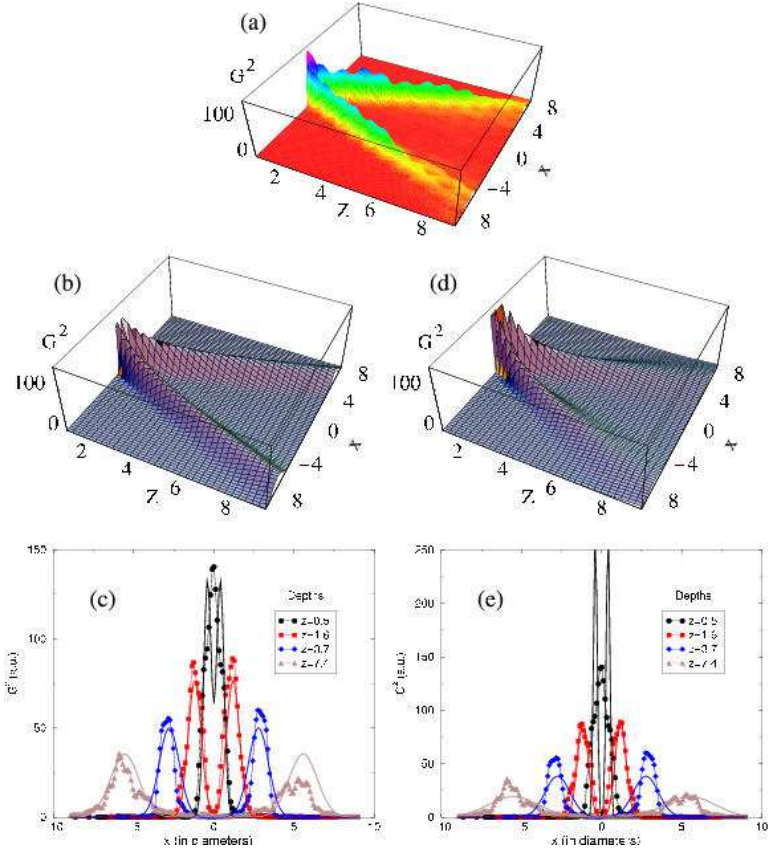}}
\caption{Comparison of nonlinear least-squares fits to the CD
equation and the CW equation for a rectangular packing of disks with a
frictional coefficient $\mu=0.48$. a) a perspective 3D plot of the
experimental response, b) a 3D plot of the least-squares fit to the CD
equation, and c) comparison of profiles of the experimental data and
the fitting data at different depths, where the symbols are
experimental data and solid lines are fits. d) and e) are similar to
b) and c), but for the CW equation. Here, x and z are measured in disk
diameters.}
\label{fig:CD_CW_comp}
\end{figure}

\hspace{0.1in} It is useful to compare the observations on bidisperse
and rectangular packings to the predictions of Claudin et
al\cite{claudin98}.  In the case of weak disorder, these authors
predict a Convection-Diffusion equation, as discussed in the Appendix.
This equation is characterized by a propagation speed, $c$, that
determines the opening angle of the two-peak response, and by a
diffusivity, $D$, that determines the rate at which the peaks broaden
with depth.  These authors predict that $D$ grows and $c$ decreases as
the disorder increases.  In order to make contact with this model, we
determined c and D by finding nonlinear least-squares fits of the mean
responses for a given type of system at all depths to the CD equation
simultaneously. At large depths, the data approach the noise floor,
and it is not possible to resolve the two-peak structure.
Accordingly, we fit only the regions from about 2 to 10 grains
deep. In Fig.~\ref{fig:CD_fit}, we show the resulting coefficients c
and D extracted from the disk data and from several other data sets
discussed below. The error bars in Fig.~\ref{fig:CD_fit} represent a
$95\%$ confidence interval for each parameter. In
Fig.~\ref{fig:CD_CW_comp}, we show an example of such fits.
Fig.~\ref{fig:CD_CW_comp}a shows a perspective 3D plot of the mean
response from experiments for a rectangular packing of disks with a
frictional coefficient $\mu=0.48$, and Fig.~\ref{fig:CD_CW_comp}b
shows the least-squares fit of the experimental data to the CD
equation. In Fig.~\ref{fig:CD_CW_comp}c, we compare for various depths
the profiles of the experimental data (symbols) and the fits to the CD
equation (solid lines). One can quantify the goodness of a nonlinear
fit by calculating a value $R^2$, a so-called coefficient of
determination\cite{fitbook}. The closer that $R^2$ is to 1, the better
the fit is. In our fits, the $R^2$ values are in the range of $0.8
\sim 0.9$.  An excellent fit corresponds to values of $R^2$ only
slightly less than 1. However, considering the complexity of the data
and fits, we believe the data are described reasonably, although not
exactly, by this model.

\begin{figure}[h]
\center{\includegraphics[width=3.75in]{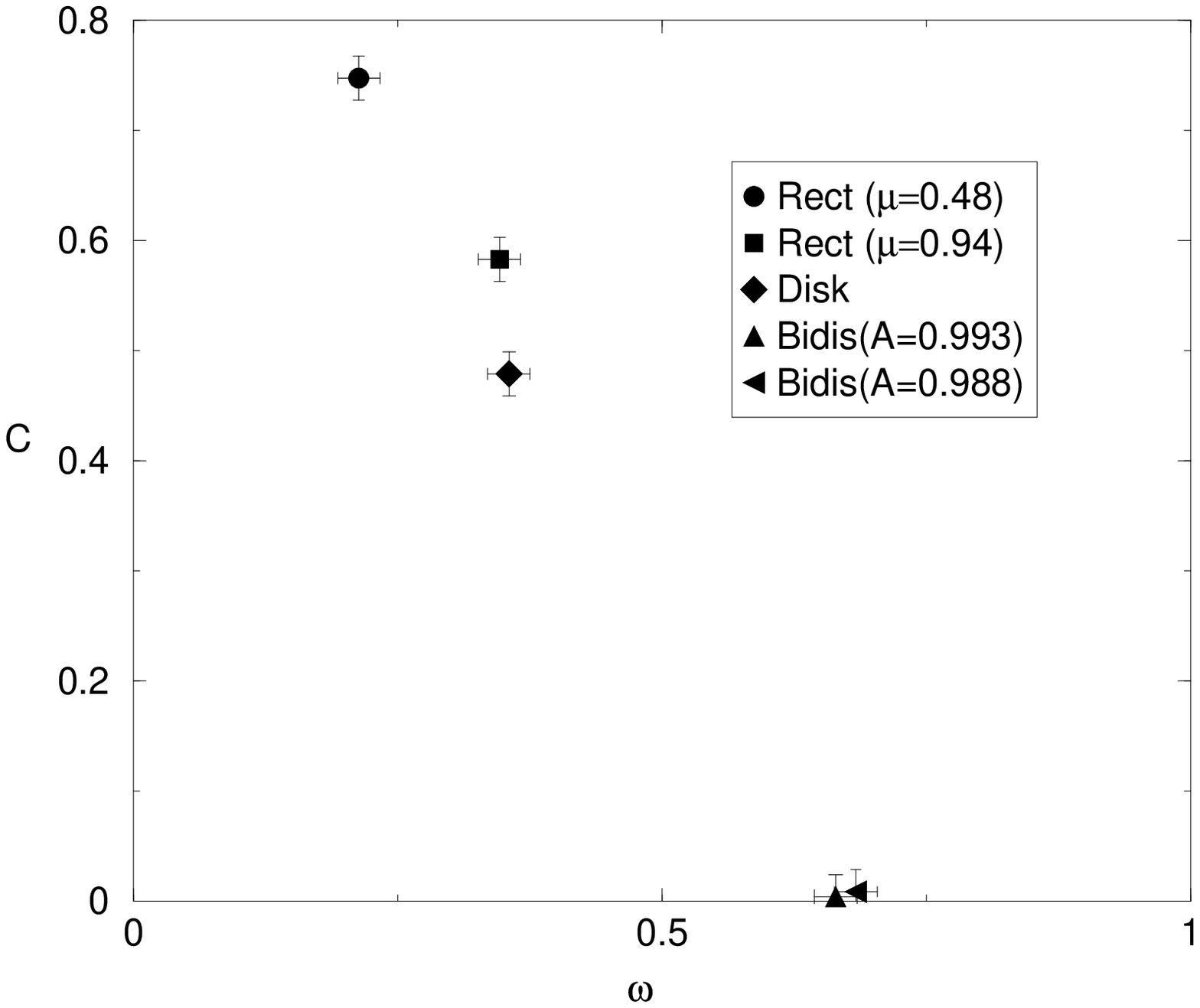}}
\caption{The coefficients of c and $\omega$ extracted by fitting the
mean responses from different systems to the CW equation. Each point
in the graph corresponds to a different system. See the text for more
detail.}
\label{fig:CW_fit}
\end{figure}

\hspace{0.1in} In total, Fig.~\ref{fig:CD_fit} shows fit results for $c$ and $D$ of
the CD equation for five different systems.  These systems include two
rectangular lattices with friction coefficients $\mu =0.48$ and $\mu
=0.94$, a hexagonal packing of monodisperse disks, and the two
randomly packing bidisperse disk systems with ${\mathcal{A}}=0.993$ and
$0.988$.  This figure suggests that, as the disorder in the system
increases, the coefficient c decreases and the coefficient D
increases.

\hspace{0.1in} One possible way to distinguish between predictions of
Bouchaud et al.\cite{doubleY,socolar_02} and anisotropic elasticity
models\cite{goldhirsch} is by determining how the width of each peak
changes with depth. For the former, one expects a width that grows
with square-root of depth, and for the latter, a width that grows
linearly with depth\cite{socolar_02}.  Note that the width for the
data of Fig.~\ref{fig:rect_width} is nearly constant at roughly a
particle diameter for these data, until a depth of about four particle
diameters, whereafter it grows with depth.  The data for the width
vs. depth then suggest more of a linear variation than a square-root
variation, particularly in the data of Fig.~\ref{fig:rect_width} that
have not been coarse-grained.  At this time, however, based on our
data, it does not seem possible to distinguish between these two
models.

\hspace{0.1in} Nevertheless, it would be useful to have a simple functional form,
similar to the solution of the CD equation (see Appendix) to which we
could fit complete data sets.  Although at this time we are not aware
of a specific (simple) functional form for force transmission in an
elastic system with disorder, we have fitted the results to a
functional form that has a number of the properties that we expect
from such a solution.  These include preferred propagation directions,
and a width that increases linearly with distance from the source.
The functional form that we have used, denoted as the CW equation, has
these properties, as discussed in the Appendix. This function consists
of two gaussians that propagate along the direction defined by a
velocity, c, and that widen linearly with depth.  That is, we replace
the width function, $W \propto z^{1/2}$ in the CD model with $W=\omega
z$, where $\omega$ is a system-dependent constant. The two parameters
in the CW equation are the propagation speed $c$ and $\omega$.  These
play similar roles to c and D in the CD equation. In
Fig.~\ref{fig:CD_CW_comp}d and e, we show a sample fit to the CW
equation.  In Fig.~\ref{fig:CW_fit}, we show the fitted coefficients c
and $\omega$, where the quality of the fits is comparable to what was
obtained with the CD equation. As the disorder in the system
increases, the coefficient c decreases and the coefficient $\omega$
increases, an effect that is similar to the results in
Fig.~\ref{fig:CD_fit}. For the two-peak behavior shown in Fig.~\ref{fig:CD_fit}
 and \ref{fig:CW_fit}, the values of c correspond rather well, within uncertainties, to the value of the lattice directions assumed by the packing geometry as in Fig.~\ref{fig:resp_rect_pics}. This is consistent with the point of view of anisotropic elasticity models\cite{goldhirsch}.

\subsection{Non-normal force responses}

\hspace{0.1in} We consider the vector character of force propagation in this section,
namely the response to forces applied at arbitrary angles to the
surface.  We first consider the response to a non-normal force in a
disordered system, one consisting of pentagonal particles.  We then
consider the corresponding problem for an ordered system, in this case
monodisperse disks in a triangular packing.  

\hspace{0.1in} As in the previous measurements, particles were placed in a vertical
plane, and forces were applied on a single grain at an angle $\theta$
with respect to the horizontal direction. Specifically, a force of 50g
was applied to the surface at angles, $90^{o}$, $60^{o}$, $45^{o}$ and
$30^{o}$ with respect to the horizontal for pentagons and at $90^{o}$,
$75^{o}$, $60^{o}$, $45^{o}$, $30^{o}$ and $15^{o}$ with respect to
the horizontal for a hexagonal packing of disks. The line of force was
chosen so that, as much as possible, it passed through the center of
gravity of the grain. In other respects, the procedure and analysis
were the same way as we described previously.

\begin{figure}[t]
\center{\includegraphics[width=3in]{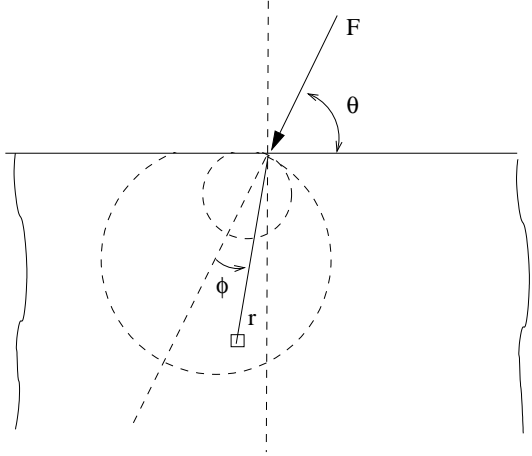}}
\caption{Schematic drawing of a force applied to the surface of an
elastic semi-infinite plate at an angle $\theta$ with respect to the
horizontal. Here, $r$ is the distance from the point of contact, and
$\phi$ is the angle from the direction of the force, measured
counter-clock-wise.}
\label{fig:elastic}
\end{figure}

\hspace{0.1in} For the case of a disordered system, such as a packing of pentagonal
particles, we have shown\cite{gengprl_01}, as a special case, that an
applied normal force at a boundary produces a response that resembles
that of an elastic solid.  As a point of reference, when a force is
applied to an elastic plate of thickness $t$ at an arbitrary angle
$\theta$ with respect to the horizontal direction, as depicted in
Fig.~\ref{fig:elastic}, the stress tensor components are\cite{landau}:
\begin{equation}
\sigma_{rr} = \frac{2Ft}{\pi r}cos\phi, \hspace{0.1in}\sigma_{r\phi} = \sigma_{\phi\phi}=0,
\label{eq:elastic}
\end{equation}
where the angle $\phi$ is measured from the direction of the applied
force, $r$ is the distance from the point under consideration to the
point of contact.  When converted to Cartesian coordinates, where the
z-axis is aligned with the applied force, the stress components have a
simple scaling form, as seen for instance in the stress component
\begin{equation}
\sigma_{zz}(x,z) = \frac{2F}{z\pi}\frac{1}{[(x/z)^2+1]^2}.
\label{eq:elastic_zz}
\end{equation}
Note that $z \sigma_{zz}(x,z)$ depends only on the ratio of $x$ to the
depth, $z$.  A similar conclusion applies to all components of the
stress tensor.

\hspace{0.1in} The dashed circles shown in
Fig.~\ref{fig:elastic} represent loci of equal
stress, $\sigma_{rr}$. For the case of
an elastic plate, when the direction of applied force changes, these
equal-stress lines remain the same with respect to the direction of
the force, except for those points that lie outside of the material.

\begin{figure}[t]
\center{\includegraphics[width=3.70in]{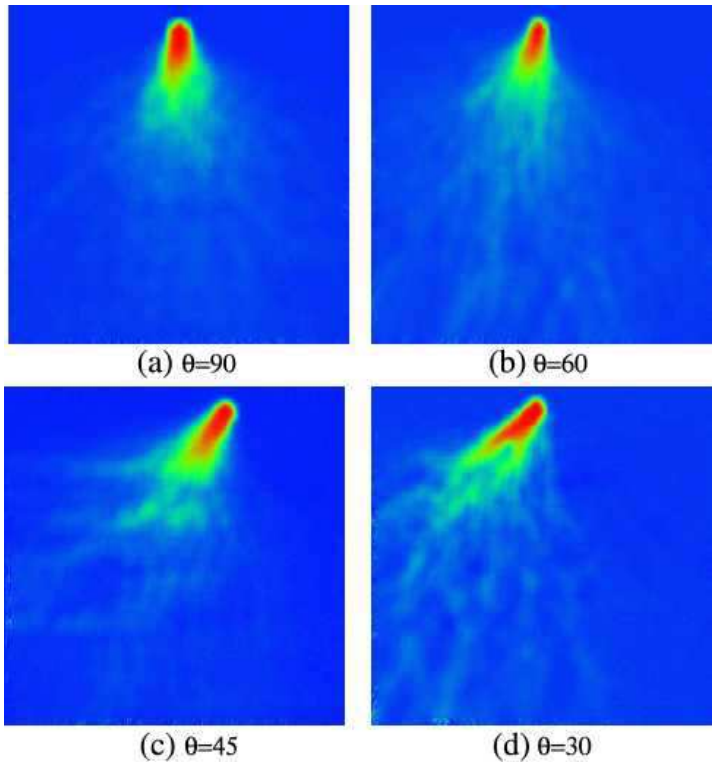}}
\caption{Non-normal force responses for a system of pentagonal
particles. A force of 0.5N was applied on the surface of the
sample. The force directions are: (a). $90^{o}$, (b) $60^{o}$, (c)
$45^{o}$ and (d) $30^{o}$, with respect to the horizontal.}
\label{fig:oblique_pent_pics}
\end{figure}

\begin{figure}[t]
\center{\includegraphics[width=4.9in]{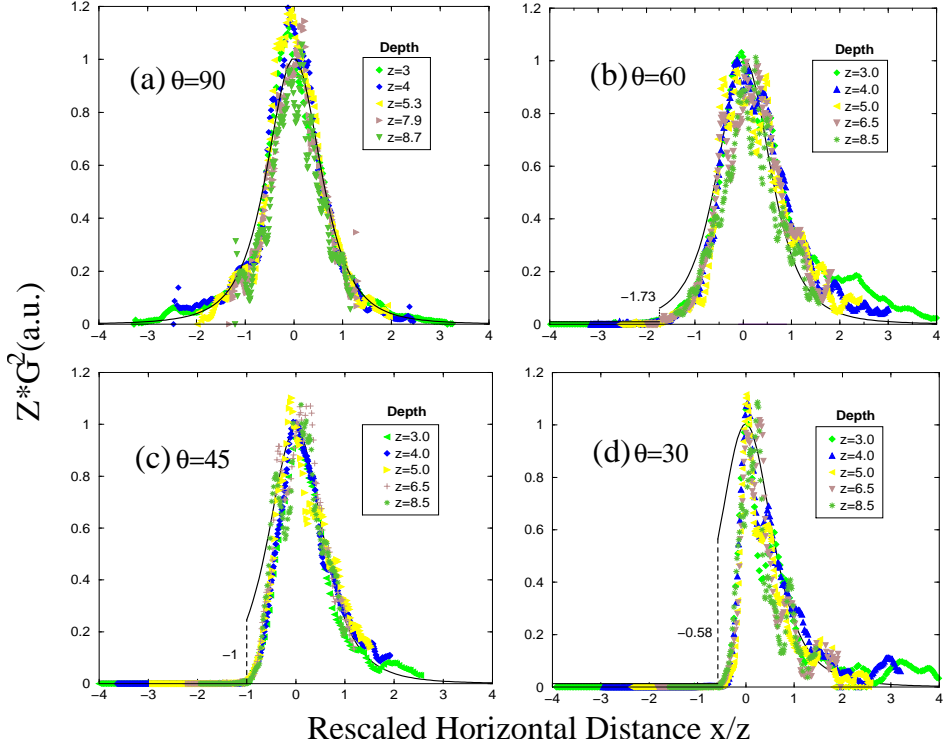}}
\caption{Rescaled mean force responses in a system of pentagonal
particles for non-normal forces applied at various angles: (a)
$90^{o}$, (b) $60^{o}$, (c) $45^{o}$ and (d) $30^{o}$. To obtain this
figure, we first rotate the coordinate axes for the response so that
the vertical axis is along the direction of the applied force, and we
then obtain the force responses along a series of horizontal lines at
depths, z, measured from the source. We then rescale these responses
as follows: 1) we normalize the $x$ coordinate by the depth, $z$ in
the rotated frame, and 2) we multiply the stresses by $z$ in the
rotated frame. Solid lines are the semi-infinite elastic plate
solution.}
\label{fig:oblique_pent_data}
\end{figure}

\hspace{0.1in} In Fig.~\ref{fig:oblique_pent_pics}, we show the grey-scale
representation of the average responses for pentagonal particles.  In
general, the force responses are centered along the direction of the
applied forces and, particularly for larger angles, they are similar
to a rotated version of the response when a normal force is
applied. This is more evident in Fig.~\ref{fig:oblique_pent_data}. To
obtain this figure, we first rotated the reference frame so that the
z-axis corresponds to the direction of the applied force.  From the
data, we computed the force responses along a series of horizontal
lines at depths, z, in the rotated coordinate system. We then rescaled
these responses as follows: we normalized the $x$ coordinate by the
value of the depth, $z$ in the rotated frame, and we multiplied the
stresses by $z$ in the rotated frame.  For comparison, we plot the
elastic plate solutions based on Eq.~\ref{eq:elastic} in these
figures. Fig.~\ref{fig:oblique_pent_data}a is a confirmation that the
response to a normal force is consistent with that of a 2D elastic
material, i.e., the widths of response vary linearly with the
depth. Fig.~\ref{fig:oblique_pent_data}b,$\,$c and d show that the
mean responses to forces at other angles in a pentagonal system have
the scaling property of an elastic medium.  However, the response
function on the side towards which the force is directed clearly
deviates from the elastic solution.  The deviations from the elastic
solution on this side in b, c and d may be attributable to the fact
that there are no tensile forces in a granular material.  Thus, this
figure suggests that on the side opposite to the direction of the
applied force contacts may open, a process that has no analogue in the
formalism of a conventional elastic solid.  It seems likely, however,
that this process is still reversible, since there was no observable
rearrangement of the particles, once the applied force was removed.

\begin{figure}[h]
\center{\includegraphics[width=5in]{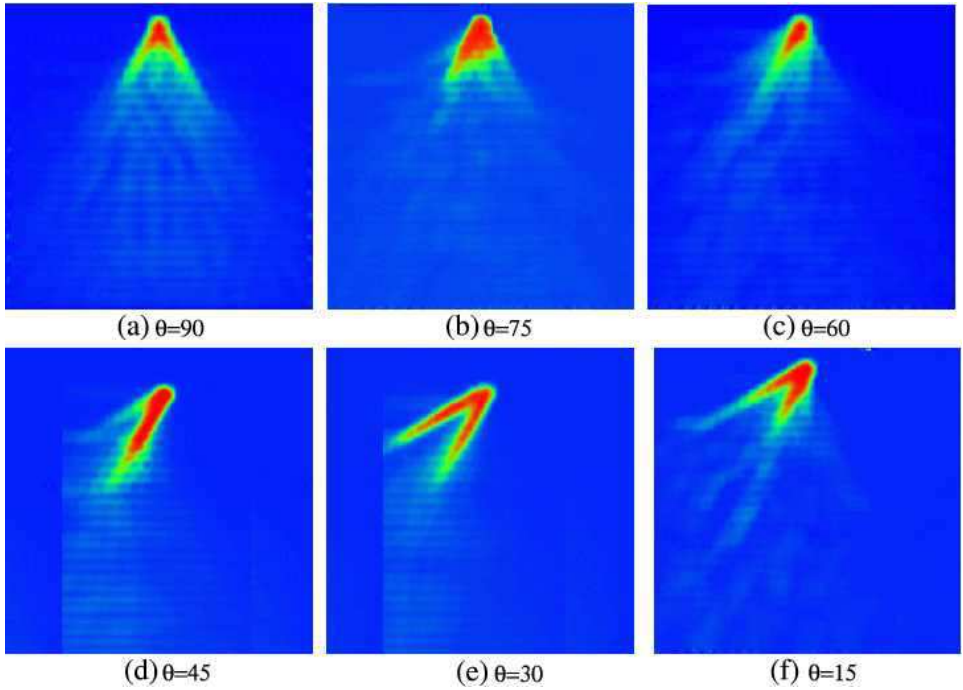}}
\caption{Responses for a non-normal force applied to a hexagonal
packing of monodisperse disks. A force of 0.5N was applied on the
surface of the sample. The force directions are: (a) $90^{o}$, (b)
$75^{o}$, (c) $60^{o}$, (d) $45^{o}$, (e) $30^{o}$ and (e) $15^{o}$,
and all angles are with respect to the horizontal.}
\label{fig:oblique_disk_pics}
\end{figure}

\begin{figure}[h]
\center{\includegraphics[width=3.75in]{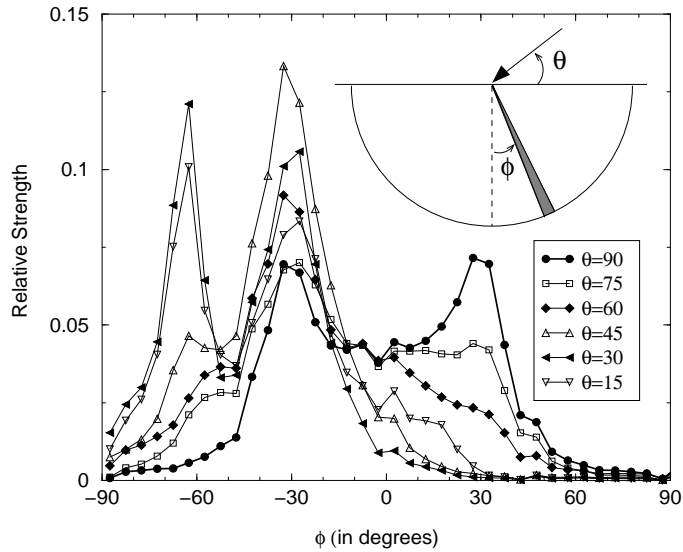}}
\caption{Relative strength of the response in a given direction, $\phi$, when a force  is applied at an angle, $\theta$. The definition of $\phi$ is illustrated in the inset.}
\label{fig:ani_strength}
\end{figure}

\hspace{0.1in} Next, we consider the response to a non-normal force on the boundary
of a triangular packing of monodisperse disks. In
Fig.~\ref{fig:oblique_disk_pics}, we show the grey-level average
response pictures for systems where a force of 50g was applied to the
surface at angles: $90^{o}$, $75^{o}$, $60^{o}$, $45^{o}$, $30^{o}$
and $15^{o}$, with respect to the horizontal. For the $\theta=90^{o}$
case, the mean response is along the two lattice directions closest to
the applied force direction, and left-right symmetry is preserved. For
the $\theta=75^{o}$ case, the left-right symmetry is broken, but the
response still involves the same two lattice directions.
Specifically, the response along the left lattice direction (i.e. most
closely aligned along the applied force direction) is strong, and the
response along the right lattice direction is relatively weak. For the
$\theta=60^{o}$ case, where the force is applied along only one of the
principal lattice directions, we see that the response is only along
that direction. For each of the $\theta=45^{o},30^{o}$ and $15^{o}$
cases, part of the average responses is aligned along the left
principal lattice direction, and part is aligned along a new direction
that is $\sim 62.5^o$ clock-wise from the vertical direction. This is
illustrated more quantitatively in Fig.~\ref{fig:ani_strength}. To
obtain this figure, we first partitioned the responses of
Fig.~\ref{fig:oblique_disk_pics} into small angular bins, $5^o$ in
width, and we then calculated the integral over the radial direction
of the responses in those bins.  Thus, these curves show the total
strength of the response in a given direction. The two principal
lattice directions are $-30^o$ and $+30^o$.  The other direction,
$\phi=-62.5$, is associated with the next-nearest neighbor lattice
direction which is aligned most closely with the direction of the
applied force.  It is interesting to note that propagation along this
direction involves a more complex process than that involved with
particles that are aligned along the principal lattice directions.  It
seems likely that friction plays an important role in this process.

\begin{figure}[h]
\center{\includegraphics[width=4.5in]{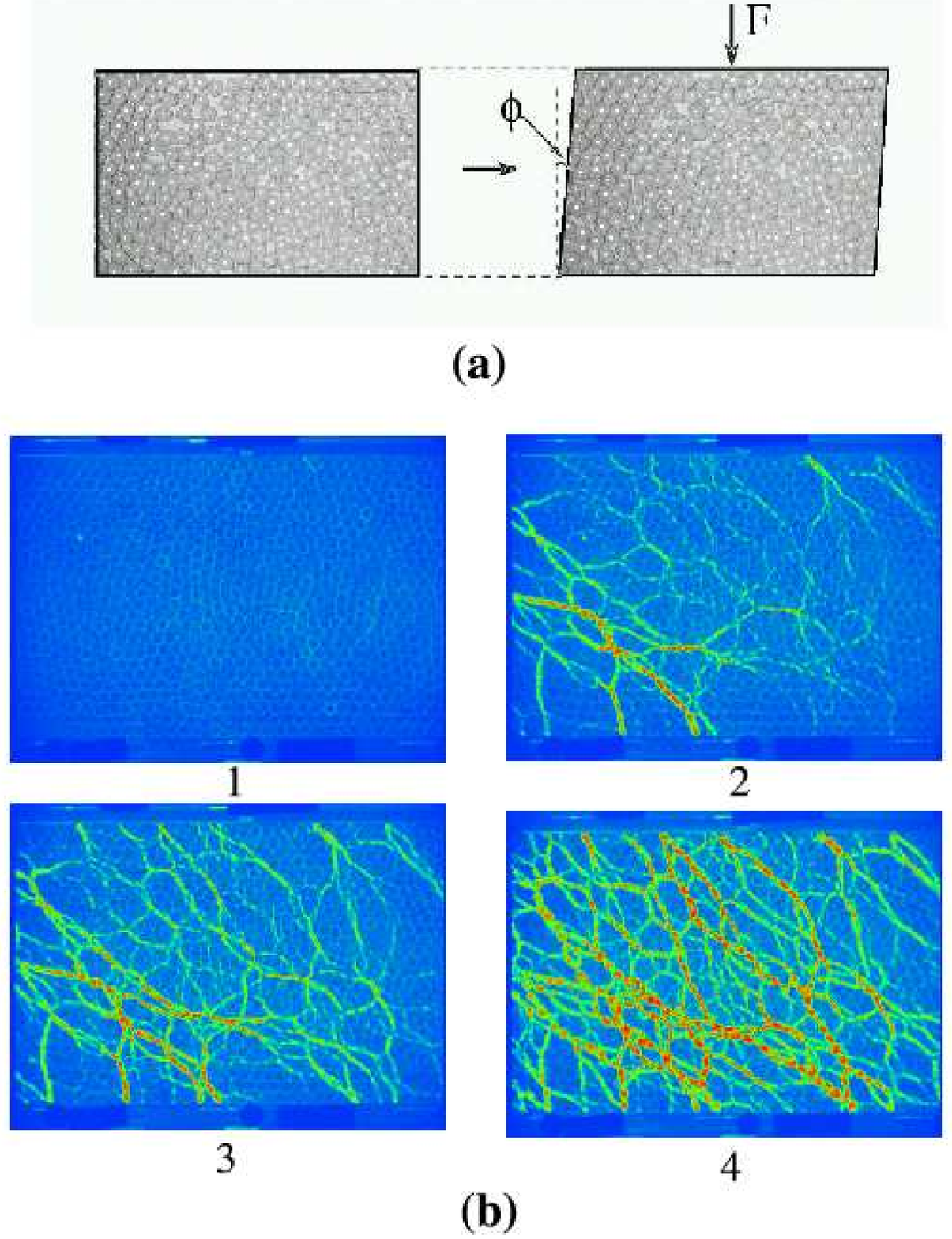}}
\caption{(a) Schematics of the 2D shearing cell with real images of
the pentagonal particles overlayed. The small black dots on each
particle denote their centers of mass.  (b). A series of photoelastic
images showing stress chain patterns for different amounts of shear
deformation. The shear deformation increases with the image
number: $\phi$=$0$, $2.4^o$, $3.2^o$ and $4.8^o$ for image 1, 2, 3 and 4, respectively.}
\label{fig:shearcell}
\end{figure}

\subsection{Responses of a system with shear deformation}

\hspace{0.1in} In this final section of experiments, we consider the characterization
of stress chain orientation and length in a system that has been
subjected to a modest amount of uniform shear, and how such systems
respond to applied external forces.  In essence, this is a probe of
the nature of anisotropic texture.

\hspace{0.1in} We created an anisotropic texture by applying nearly uniform simple
shear.  To do so, we constructed an experimental setup that is
sketched in Fig.~\ref{fig:shearcell}(a).  The particles, in this case
pentagons, rested on a flat horizontal surface consisting of
Plexiglas, and they were confined by Plexiglas walls.  Two parallel
boundaries were hinged at their lower corners so that it was possible
to shear the system.  The other two boundaries remained parallel
during the shearing process.  One of these latter boundaries remained
fixed relative to the Plexiglas bottom plate, and the other, which was
opposite the hinges, was guided so as to keep constant the distance
between the opposite parallel boundary.  Hence the available area to
the particles remained constant.  The system size was about $\sim
47\,{\rm cm} \times \sim 22\,{\rm cm}$.  We applied controlled amounts
of shear to this system by slowly displacing the upper left corner of
the boundary by a measured amount.  For this experiment, we used 1167
pentagonal particles that were $\sim 6.3\,{\rm mm}$ on an edge.  The
packing fraction was 0.795.

\hspace{0.1in} The experimental procedure was the following: (1) the left edge was
pushed continuously until it reached a given angle, $\phi$, with
respect to the normal direction. A typical series of stress patterns
as $\phi$ was increased is shown in Fig.~\ref{fig:shearcell}(b). (2)
Then a small local force perpendicular to the top edge was applied on
a particle at the top boundary and the response image was
recorded. (3) The force was removed and the background image was
recorded. (4) We then followed the general image processing procedure
as described above to obtain the mean response.  By taking images
without polarizers, we were able to obtain particle positions, as
shown in Fig.~\ref{fig:shearcell}(a), and thus, we were able to
calculate the texture tensor. We also characterized the orientation
and length of stress chains which are typified by
Fig.~\ref{fig:shearcell}(b), through the use of two point spatial
correlation functions for the stress.

\begin{figure}[h]
\center{\includegraphics[width=3.5in]{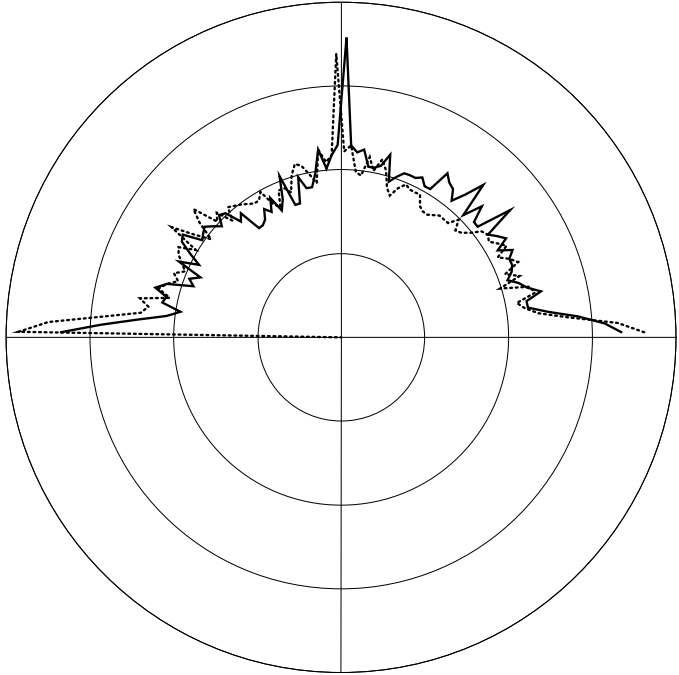}}
\caption{Comparison of distributions, $\rho(\theta)$, of fabric tensor principal directions before (Dotted line) and after (Solid line) the shear deformation is applied. The distributions are averaged over all the particles and over 50 runs.}
\label{fig:fabric_shear}
\end{figure}

\hspace{0.1in} We first examine the impact on the texture, characterized by the
fabric tensor, as in Eq.~\ref{eq:fabric}, due to the shear deformation.
Since there can be a range of distances for two pentagons to be in contact, we consider a contact to exist if the distance between centers of two 
pentagons falls within the interval [$r_{min}$, $r_{max}$], where $r_{min}$ and $r_{max}$ correspond to the minimal and maximal distance between centers of two pentagons while they are still in contact. Thus, the contacts between pentagons may be overestimated. 
When the fabric tensor is diagonalized, its major principal direction,
$\theta$, gives the direction, along which the particles have, on
average, the largest number of contacts. A comparison of the
distributions of these principal directions, $\rho(\theta)$, for all
particles before and after the application of a shear deformation to
the system, provides a quantitative measure of how much the geometric
contact structure has changed. In Fig.~\ref{fig:fabric_shear}, we show
such a comparison. The dotted line shows the distribution before the
shear deformation is applied and the solid line is after the shear
deformation. These data are averaged over all the particles and over
50 runs. We observe that there are more contacts along the horizontal
and vertical directions than any other directions for both the
``before'' and ``after'' cases, which can be explained by the fact
that the particles align with the boundaries. In the ``after'' case,
the distribution is slightly skewed from that of the ``before''
case. Notably, the change in texture is not nearly as significant as
the changes in the stress chains, which we will discuss next. However, caution needs to be taken in interpreting data show in Fig.~\ref{fig:fabric_shear}. Since a relative small change in the displacement, in the order of microns, is enough to produce a large contact force, those minute structural changes may not be detected in the current experimental measurements using fabric tensors, which has a resolution of about 0.5 mm (1 pixel).  


\hspace{0.1in} In Fig.~\ref{fig:shearcell}b, we show the impact on the stress chains
caused by applying a small shear deformation.  This set of images
follows the course of a deformation beginning with $\phi = 0^o$ and
ending with $\phi = 4.8^o$.  An obvious result of this deformation is
that the stress chains tend to align in a direction that opposes the
deformation, but at an angle that greatly exceeds the angular strain.
A similar stress chain alignment was observed in 2D Couette
shear\cite{howell_99}, although in this case the strains were very
large. In the present experiments, the stress chain orientation tended
to saturate following a small angular deformation, i.e. for $\phi
\stackrel{>}{\sim} 5^o$, the typical stress chain angle did not
significantly change.  We return to this point below.

\hspace{0.1in} An important issue concerns the spatial structure of the stress chains
that are generated in response to such a deformation.  With images
such as those shown in Fig.~\ref{fig:shearcell}(b)3 and 4, where
stress chains are well defined, we can characterize the stress chain
orientation and chain length by calculating the spatial
auto-correlation $c(\bf{r})$ for the stress, i.e. $G^2$:
\begin{equation}
c(\textbf{r})=c(r,\theta)=<G^2(\textbf{x})G^2(\textbf{x}+\textbf{r})>,
\label{eq:correlation}
\end{equation}
where the brackets denote an average over spatial coordinates
$\textbf{x}$. In this calculation, it is important to retain angular
information in $c(r,\theta)$ in order to extract information about the
anisotropic features of the system.  The actual calculation of the
correlation function is performed in wavenumber space using FFT (Fast
Fourier Transform) techniques, since the computation in the space
domain is cumbersome when the image size is large, e.g. 512 $\times$
512 pixels.

\begin{figure}[b]
\center{\includegraphics[width=5in]{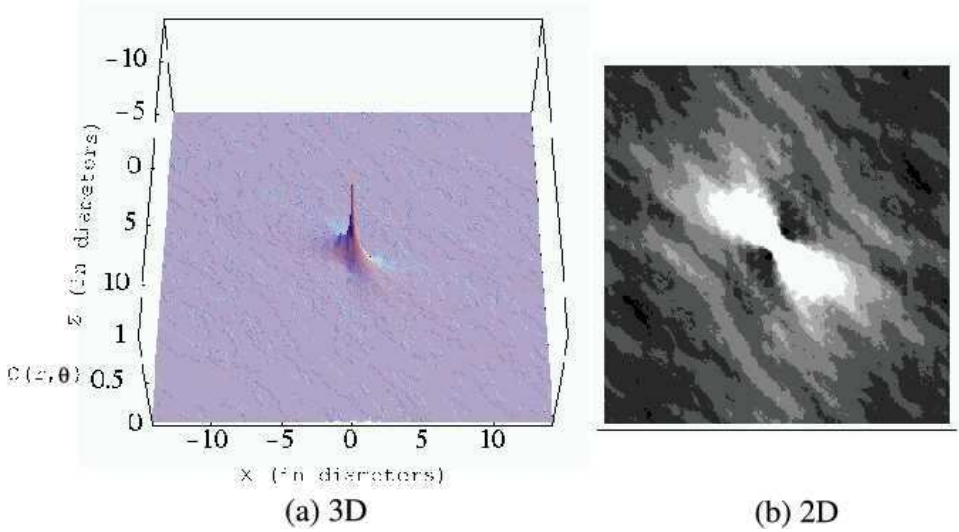}}
\caption{3D (a) and 2D (b) representations of the spatial
auto-correlation function $c(r,\theta)$ for stresses in a shear cell
(as typified by Fig.~\ref{fig:shearcell}(b)3,4).  These data are an
average of 50 independent realizations.  The image size is 512
$\times$ 512 pixels, cropped from the original image which is 640
$\times$ 480 pixels and padded at the edge with the mean intensity,
for computational efficiency. These images show that the correlation
along the stress chain direction is much stronger than along the
perpendicular direction.}
\label{fig:correl_2D_3D}
\end{figure}

\begin{figure}[b]
\center{\includegraphics[width=4.75in]{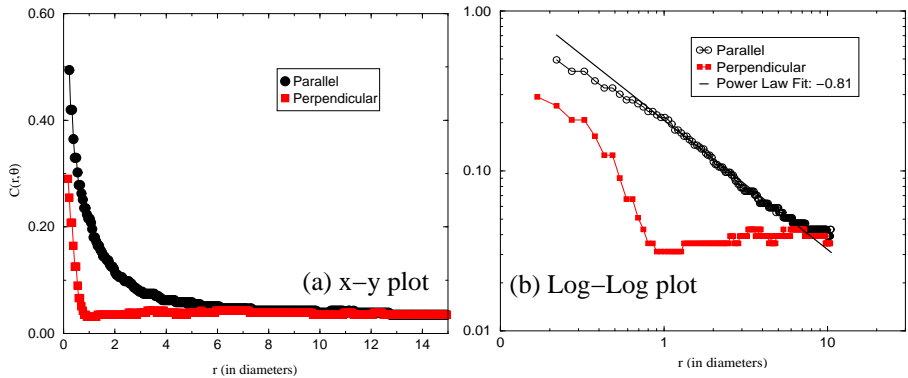}}
\caption{(a) Spatial auto-correlation function $c(r,\theta=135)$
(parallel to the stress chain direction) and $c(r,\theta=45)$
(perpendicular to the stress chain direction). Here, $\theta$ is
measured from the right horizontal direction. (b) same data shown on
double logarithmic scales.  The correlation function parallel to the
stress chain direction can be fitted with a power law: $c(r, \theta)
\sim r^{-\gamma}$, where $\gamma$ is 0.81, showing a persistent long
range order.  In the direction transverse to the chains, the
correlation function falls to the background value over a length that
is roughly one grain size.  Note that distances are measured in
particle sizes.}
\label{fig:correl_powerlaw}
\end{figure}

\hspace{0.1in} Fig.~\ref{fig:correl_2D_3D} shows such a spatial auto-correlation
function $c(r,\theta)$ in a perspective 3D plot on the left, and in
greyscale on the right. These data are obtained by averaging 50
realizations. Clearly, and perhaps not surprisingly, these images show
that correlation along the stress chain directions is much longer
range than along the perpendicular direction, even though the stress
chain directions span a finite range of angles.  The strongest
direction for $c(r,\theta)$ is $45^{o}$ from the
vertical. Fig.~\ref{fig:correl_powerlaw} shows the correlation
function evaluated along this direction and the direction
perpendicular to it. Along the perpendicular direction, the
correlation is almost a $\delta$ function, dropping rapidly to a value
close to zero over a distance of about 1 grain diameter. However,
along the strong direction, the correlation function is consistent
with a power law with an exponent of $-0.81$, showing long range order
over the size of the system.

\hspace{0.1in} These data may also shed some light on an apparent
conflict between different force measurements by Liu et
al.\cite{coppersmith} and by Miller et al.\cite{miller_96}.  In the
first set of experiments, a granular system was subject to uniaxial
compression, and forces were then measured in a plane that was normal
to the direction of compression.  No correlations were observed
between forces in the plane of the measurement.  In the second set of
measurements, stresses were measured at the boundaries of a system
undergoing plane shear.  In this case, the data indicated correlations
in the forces.  The possible explanation here may be that for the
first experiment, the stress chains were normal to the plane of
measurement, whereas in the latter, the stress chains were likely
tilted relative to the plan of measurement by something like $45^o$.

\begin{figure}[b]
\center{\includegraphics[width=3.0in]{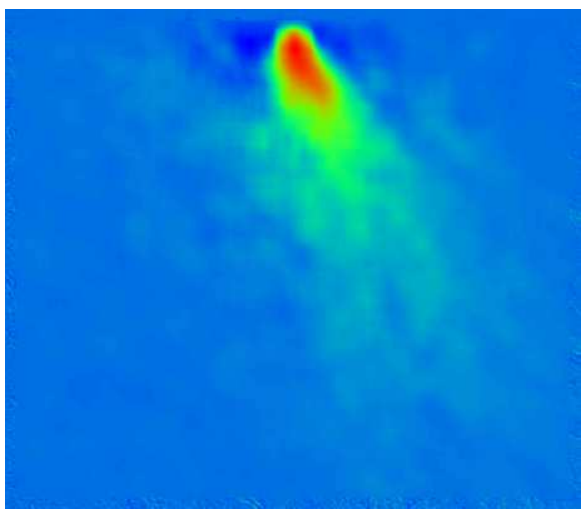}}
\caption{Mean force response in a pentagonal system that was prepared
with a shear deformation of $4.7^{o}$. The force is applied normal to
the top edge. Note the sheared-induced anisotropic contact network has
significantly changed the force propagation direction, in contrast to
the response for an isotropic system of pentagons.}
\label{fig:shear_resp_pics}
\end{figure}

\hspace{0.1in} The angle $45^o$ seen in the correlation function above,
can be understood by noting that for small angles of
shearing, $\phi$, simple shear can be expressed as a solid-body rotation by
$\phi /2$ plus compression along a line oriented at $45^o$
(the strong stress chain direction) and an expansion at $90^o$
to that direction.  Thus, the strong asymmetry set up in the stress
network is associated with strengthening of contacts due to the
compression.  It seems likely that this strengthening can occur
in the case of angular non-space-filling particles with minimal
change in the contacts, and hence in the texture.

\begin{figure}[]
\center{\includegraphics[width=5.5in]{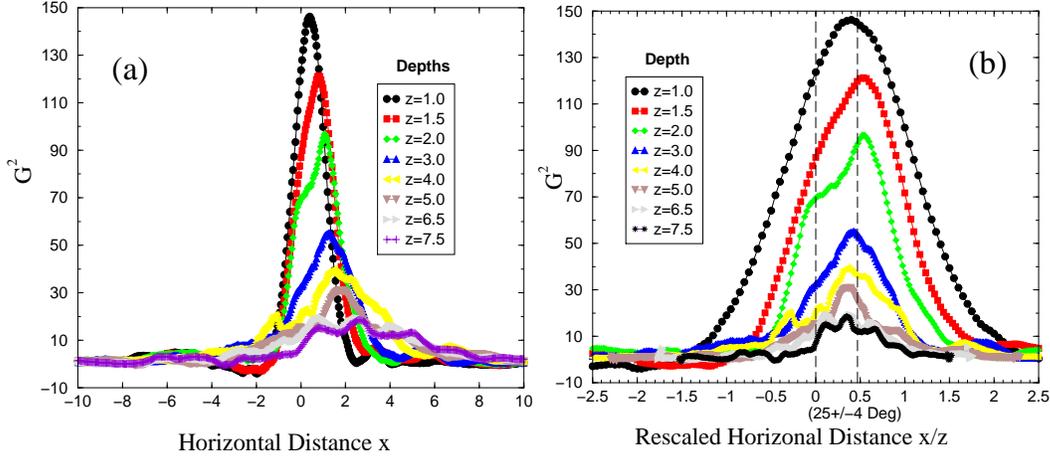}}
\caption{Quantitative representation of the data from the previous
figure for a pentagonal system with shear deformation of $4.7^{o}$:
(a) The averaged photoelastic response, $G^2$, v.s. horizontal
distance, x, at various depths z. (b) The same data as (a), but with
the x coordinate rescaled with the depth z. In both plots, x and z are
measured in grain sizes, where a grain size is about 1 cm.}
\label{fig:shear_resp_data}
\end{figure}

\hspace{0.1in} Lastly, we consider the impact of anisotropy in the stress chain
network on force propagation. When a vertical force was applied, the
force tended to propagate along or toward the stress chains.
Specifically, on average, a force applied normally to the upper
boundary led to a force response whose peak deviated from the vertical
direction, as shown in Fig.~\ref{fig:shear_resp_pics}. In
Fig.~\ref{fig:shear_resp_data}a, we give data for the response at
different depths for a $4.7^o$ deformation. In
Fig.~\ref{fig:shear_resp_data}b, we plot the same data but we rescale
the x coordinates with depth z. In this latter figure, all peaks of
the responses at different depths are roughly located around
$x/z=0.5$, which is about $25\pm 4^{o}$ from the vertical.

\section{Summary and conclusions}

\hspace{0.1in} We have measured the force response of 2D granular systems to local
perturbations under various conditions. There are large variations
from realization to realization, and we consider ensembles built from
many repeated observations under identical conditions. We have
obtained ensemble-averaged responses for five types of systems:
monodisperse systems packed in an ordered triangular lattice,
bidisperse systems with different amount of disorder, monodisperse
systems packed in an ordered rectangular lattice, systems with forces
applied at an arbitrary angle at the surface, and systems that have
been subject to shear deformation, hence with textured/anisotropic
features. We find that disorder, packing structure, friction and
textures affect the average force response in a granular system
significantly. Specifically, we have found that: 1) in ordered
triangular packings, normally applied forces propagate along the
principal basis vectors of the lattice; 2) in bidisperse systems, when
the amount of disorder is increased by adjusting the size and number
ratio of large and small disks, the average response to normally
applied forces changes from a response with a two-peak feature to a
one-peak response; 3) in a rectangular lattice system, forces
propagate along the lattice directions and when the friction between
particles is decreased, the mean response becomes sharper; 4) when a
force of arbitrary direction is applied at the surface of a disordered
packing (pentagonal particles) the mean response can be described by
an elastic solution; 5) when a force of arbitrary direction is applied
at the surface of an ordered packing (hexagonal packing of disks) the
mean response propagates along lattice directions which may include
next-nearest-neighbor directions; 6) in a system with shear-induced
anisotropy, the stress chains tend to orient along roughly a $45^o$
degree angle so as to strongly resist the additional deformation; 7)
in such an anisotropic system, force correlation along the preferred
direction is long-range, and the correlation function is a power law
with an exponent of $-0.81$; 8) and in such an anisotropic system, the
resulting average response to a normal local force tends to propagate
along or toward the preferred force direction.

\hspace{0.1in} These results help identify the important factors that affect force
propagation in granular media and thus raise the need to incorporate
these factors into models. The data are inconsistent with the scalar
q-model.  At this time, it is not possible to distinguish between
essentially propagative models by Bouchaud et al. and anisotropic
elasticity models as suggested by Goldenberg and Goldhirsch.  However,
the wave-like propagation seen by Tkachenko and Witten for
polydisperse frictionless particles was not seen in these experiments.
There are a number of important issues to address in the future.  One
of these involves improved techniques and tests to help further narrow
down the range of prospective models.  This might involve examining the
role played by the boundaries, which differs between hyperbolic and
elliptic systems.  In addition, it would clearly be valuable to
further develop the photoelastic resolution.  Other interesting and
related issues concern the force response to distributed loads and the
force response when plastic deformations occur.  We will address these
issue in future work.

\ack

\hspace{0.1in} We appreciate helpful interactions with P. Claudin, S. Luding,
C. Goldenberg, I. Goldhirsch, and J. Socolar. The work of GR and EC
was supported by PICS-CNRS $\#563$.  The work of JG and RPB was
supported by the US National Science Foundation under Grant
DMR-0137119, and DMS-0204677, and by NASA under Grant NAG3-2372.

\appendix

\section{Appendix}

\hspace{0.1in} This appendix serves two roles.  First, it summarizes some important
properties of the Convection-Diffusion equation, and of the elastic
response function for a semi-infinite solid sheet\cite{landau}.  In
addition, it uses these models to provide a setting in which to
examine the linearity of the response seen in the experiments.

\subsection{Summary of the Convection-Diffusion equation properties} 

\hspace{0.1in}  Here, we briefly review the two branch Convection-Diffusion equation(CD)
adapted from Ref.\cite{claudin98} which is intended to be relevant
for weakly disordered systems.

The CD equation is:
\begin{equation}
{\mathcal{O}^+ \mathcal{O}}^- \sigma_{zz}(x,z) = 0 ~.
\label{eq:operator}
\end{equation}    
Here, ${\mathcal{O}}^{\pm} = \partial_{z} - D \partial_{xx} \pm c
\partial_{x}$, x and z are horizontal and downward coordinates, and c
and D are two parameters in analogy to a dimensionless velocity and a
diffusion coefficient. The solution to this equation for a delta
function initial condition, $\sigma_{zz}(x,0) = F\delta(x,0)$ is
\begin{equation}
\sigma_{zz}(x,z)=\frac{F}{2}(\frac{1}{\sqrt{4\pi Dz}}
                e^{-\frac{(x-cz)^{2}}{4Dz}}+\frac{1}{\sqrt{4\pi Dz}}
                e^{-\frac{(x+cz)^{2}}{4Dz}})
\end{equation}
where $F$ is the magnitude of the downward delta function stress.

\hspace{0.1in} In order to provide an alternative description to the CD equation, we
used what we call {\bf CW equation}. Here, the point is to incorporate
some of the expected features of an elastic model into a simple
fitting function.  In a similar spirit to the CD equation, we write
\begin{equation}
\sigma_{zz}(x,z)=\frac{F}{2}(\frac{1}{\sqrt{2\pi}W}
                e^{-\frac{(x-cz)^{2}}{2W^2}}+\frac{1}{\sqrt{2\pi}W}
                e^{-\frac{(x+cz)^2}{2W^2}})
\end{equation}
where $W(z)$ is the width of each peak. For the CD model,
$W=\sqrt{2Dz}$. If we replace the diffusively increasing width by a
linearly increasing width, $W=\omega z$, where $\omega$ is a constant,
we obtain the CW description:
\begin{equation}
\sigma_{zz}(x,z)=\frac{F}{2}(\frac{1}{\sqrt{2\pi}\omega z}
                e^{-\frac{(x-cz)^{2}}{2\omega^2z^2}}+\frac{1}{\sqrt{2\pi}\omega z}
                e^{-\frac{(x+cz)^{2}}{2\omega^2z^2}}).
\end{equation}

\subsection{Elastic response and tests for linearity in ordered systems} 

\hspace{0.1in} This model provides a convenient setting to test for linearity in an
ordered disk packing.  Specifically, to carry out such a test, we
sampled points along the lattice directions of a monodisperse
hexagonal disk packing, which corresponds to the directions $x = \pm
cz$ where the peaks of responses are located. In the CD model, the
stresses at those points are only determined by the depth z, since
$\sigma_{zz}(x\pm cz)=\frac{F}{4\sqrt{\pi Dz}}e^{-\frac{c^2}{D}z}$. In
Fig.~\ref{fig:lin_test}(a), we plot the measured stress versus applied
point force at different depths. We find that when the applied force
is below about 0.5 N, there is a good linearity within the error
bars. In our experiments, in order to increase the signal to noise
ratio, we usually chose a working force close to the upper bound of
this linear region.

\subsection{Tests for linearity in disordered systems} 

\hspace{0.1in} The second model that is relevant here is the response of 
a semi-infinite elastic plate that
is subject to an applied point force at the free surface, as sketched
in Fig.~\ref{fig:elastic}.  The stresses in this case are give by
Eqs~\ref{eq:elastic} and~\ref{eq:elastic_zz}.  In particular,
Eq.~\ref{eq:elastic_zz} indicates a simple relationship for
$\sigma_{zz}$ that applies equally well to the other stress components
in Cartesian coordinates: $\sigma_{zz}(0,z) = \frac{2F}{z\pi}$.  Thus,
we expect that $z \sigma_{zz}$ should depend linearly on $F$, for all
depths. In Fig.~\ref{fig:lin_test}(b), we plot the measured stress
multiplied by the corresponding depth against the applied force. As
expected, we see that lines for different depths collapse on the same
line.  However, this linearity only holds when the applied force is
less than 0.5 N.

\end{document}